\documentclass[prl,
               preprint,
               superscriptaddress,
               onecolumn,
               showpacs,
               longbibliography,
               ]{revtex4-1}
               
\usepackage{graphicx}   
\usepackage{dcolumn}    
\usepackage{bm}         
\usepackage{amsmath}
\usepackage{amssymb}
\usepackage[usenames]{color}      

\newcommand{\Pra}{\mbox{\textit{Pr}}}  
\newcommand{\Ray}{\mbox{\textit{Ra}}}  
\newcommand{\Nus}{\mbox{\textit{Nu}}}  
\newcommand{\Rey}{\mbox{\textit{Re}}}  
\newcommand{\Lew}{\mbox{\textit{Le}}}  

\begin{document}

\title{Scaling laws and flow structures of double diffusive convection in the finger regime}

\author{Yantao Yang}
\email[Corresponds to:]{yantao.yang@utwente.nl}
\affiliation{Physics of Fluids Group, MESA+ Research Institute, and J. M. Burgers Centre for Fluid Dynamics, University of Twente, PO Box 217, 7500 AE Enschede, The Netherlands.}

\author{Roberto Verzicco}
\affiliation{Physics of Fluids Group, MESA+ Research Institute, and J. M. Burgers Centre for Fluid Dynamics, University of Twente, PO Box 217, 7500 AE Enschede, The Netherlands.}
\affiliation{Dipartimento di Ingegneria Industriale, University of Rome ``Tor Vergata'', Via del Politecnico 1, Roma 00133, Italy}
                    
\author{Detlef Lohse}
\affiliation{Physics of Fluids Group, MESA+ Research Institute, and J. M. Burgers Centre for Fluid Dynamics, University of Twente, PO Box 217, 7500 AE Enschede, The Netherlands.}
\affiliation{Max-Planck Institute for Dynamics and Self-Organization, Am Fassberg 17, 37077 G\"{o}ttingen, Germany.}

\date{\today}

\begin{abstract}
Direct numerical simulations are conducted for double diffusive convection (DDC) bounded by two parallel plates. The Prandtl numbers, i.e. the ratios between the viscosity and the molecular diffusivities of scalars, are similar to the values of seawater. The DDC flow is driven by an unstable salinity difference (here across the two plates) and stabilized at the same time by a temperature difference. For these conditions the flow can be in the finger regime. We develop scaling laws for three key response parameters of the system: The non-dimensional salinity flux $\Nus_S$ mainly depends on the salinity Rayleigh number $\Ray_S$, which measures the strength of the salinity difference, and exhibits a very weak dependence on the density ratio $\Lambda$, which is the ratio of the buoyancy forces induced by two scalar differences. The non-dimensional flow velocity $\Rey$ and the non-dimensional heat flux $\Nus_T$ are dependent on both $\Ray_S$ and $\Lambda$. However, the rescaled Reynolds number $\Rey\Lambda^{\alpha^{\rm eff}_u}$ and the rescaled convective heat flux $(\Nus_T-1)\Lambda^{\alpha^{\rm eff}_T}$ depend only on $\Ray_S$. The two exponents are dependent on the fluid properties and are determined from the numerical results as $\alpha^{\rm eff}_u=0.25\pm0.02$ and $\alpha^{\rm eff}_T=0.75\pm0.03$. Moreover, the behaviors of $\Nus_S$ and $\Rey\Lambda^{\alpha^{\rm eff}_u}$ agree with the predictions of the Grossmann-Lohse theory which was originally developed for the Rayleigh-B\'{e}nard flow. The non-dimensional salt-finger width and the thickness of the velocity boundary layers, after being rescaled by $\Lambda^{\alpha^{\rm eff}_u/2}$, collapse and obey a similar power-law scaling relation with $\Ray_S$. When $\Ray_S$ is large enough, salt fingers do not extend from one plate to the other and horizontal zonal flows emerge in the bulk region. We then show that the current scaling strategy can be successfully applied to the experimental results of a heat-copper-ion system~(Hage and Tilgner, \emph{Phys. Fluids}, \textbf{22}, 076603, 2010). The fluid has different properties and the exponent $\alpha^{\rm eff}_u$ takes a different value $0.54\pm0.10$. 
\end{abstract}

\maketitle

\section{Introduction}

Double diffusive convection (DDC) is the convection flow where fluid density depends on two scalar fields. One of the most relevant application is oceanic mixing, in which the two scalars are temperature and salinity. Temperature diffuses about 100 times faster than salinity, and this huge difference in the molecular diffusivities allows for very rich dynamics in oceanic DDC flow~\citep{Schmitt1994}. For instance, an instability can occur even when the fluid is overall stably stratified~\citep{Stern1960}. A comprehensive review of the field can be found in the recent book by~\citet{Radko2013}. 

One of the interesting phenomena in DDC is the salt finger structure, i.e. narrow elongated vertical flows which were observed in many experiments \citep[e.g. by][]{Turner1967,Linden1978,Taylor1989}. For a DDC flow bounded by two reservoirs with fixed values for the two scalars, \cite{Krishnamurti2003,Krishnamurti2009} observed either a single layer of salt fingers, or an alternating stack of salt-finger and convective layers which resembles the thermohaline staircase observed in the ocean~\citep[e.g.][]{Tait1971,Schmitt2005}. Three-dimensional (3D) direct numerical simulations (DNS) in a fully periodic domain produced layered structures which were very similar to the thermohaline staircase~\citep{Stellmach2011,Radko2014}. Numerical studies also revealed that internal gravity waves can spontaneously develop through collective instability and modulate the salt-finger field~\citep{Stern2001,Stellmach2011}. In recent DDC experiments by using electrodeposition cells, which are very close to the Rayleigh-B\'{e}nard (RB) setup, Tilgner and coworkers discovered that salt fingers also occur when the fluid is unstably stratified~\citep{Hage_Tilgner2010,Kellner_Tilgner2014}. This finding is very surprising and unexpected since one would assume that RB convection dominates in an unstably stratified system. However, linear instability analysis revealed that, in this new regime, salt-finger solutions can indeed be obtained~\citep{Schmitt2011}. Our previous numerical results with the same flow setup well agreed with those experiments~\citep{ddcjfm2015}.

Numerous theories and models have been developed in the past to understand the scalar fluxes produced by the salt-finger structures, e.g.~see~\cite{Kunze2003} and the references therein. Recently, new models were proposed for the fully periodic DDC flow and tested against numerical results, such as the mean-field theory of~\cite{Traxler2011} and the equilibrium model of~\cite{Radko2012}. For the finger layer bounded by two solid boundaries, we discovered that the Grossmann-Lohse (GL) theory, which was originally developed for RB flow and showed great success~\citep{GL2000, GL2001, GL2002, GL2004, Ahlers2009, GLrefit2013}, can be directly applied to DDC flow and accurately predicts the salinity flux for both our numerical results~\citep{ddcjfm2015} and the experimental results~\citep{Hage_Tilgner2010}.

Although we demonstrated the success of the GL theory in accurately predicting the salinity transfer rate, the behaviour of the flow velocity was not fully investigated in the previous work~\citep{ddcjfm2015}, partially due to the limited control parameters simulated therein. Here, with the help of more systematic simulations which cover a much wider parameter range, we will establish a complete description of the scaling laws for three important global responses, i.e.~the salinity and heat transfer rates and the flow velocity. Moreover, we will also investigate the flow structures, such as the salt fingers and the boundary layers. The characteristic length scales of those structures and their dependences on the flow parameters will be discussed. 

The paper is organised as follows. We first briefly describe the numerical methods and the parameter space explored (section~2). In section~3 we develop the scaling laws for the system responses. In section~4 the flow structures and further scaling relations are presented. The applicability of the new proposed scaling strategy to other fluid system is discussed in section~5. The paper ends with conclusions and an outlook (section~6).

\section{Governing equations and numerical simulations}

We consider incompressible flow with two different scalar fields and employ the Oberbeck-Boussinesq approximation, which assumes that the fluid density depends linearly on both scalars, i.e.~$\rho(\theta, s) = \rho_0 [1 - \beta_T \theta + \beta_S s]$. Here $\rho$ is the density with some reference value $\rho_0$, and $\theta$ and $s$ are the temperature and salinity relative to some reference values, respectively. $\beta$ is the positive expansion coefficient. Hereafter the subscript $\zeta=T$ or $S$ denotes the quantity associated to the scalar $\zeta$. The governing equations read
\begin{subequations} \label{eq:ddc}
\begin{eqnarray}
  \partial_t u_i + u_j \partial_j u_i &=& 
       - \partial_i p + \nu \partial_j^2 u_i + g \delta_{i3} (\beta_T \theta - \beta_S s), \label{eq:mom}  \\
  \partial_t \theta + u_j \partial_j \theta &=& \kappa_T \partial_j^2 \theta,  \\
  \partial_t s + u_j \partial_j s &=& \kappa_S \partial_j^2 s, \label{eq:sal}
\end{eqnarray}
\end{subequations}
in which $u_i$ with $i=1,2,3$ are three velocity components, $p$ is the kinematic pressure, $\nu$ is the kinematic viscosity, $g$ is the gravitational acceleration, and $\kappa_\zeta$ are diffusivities of the respective scalar components. The continuity equation is $\partial_i u_i = 0$. The fluid is vertically bounded by two parallel plates separated by a distance $L$. At the two horizontal plates no-slip boundary conditions are applied to velocity, and both scalars have fixed values. In the horizontal directions we choose a domain size much larger then the size of the salt fingers, which allows us to apply periodic boundary conditions. 

The flow is driven by the scalar differences between the two plates. We define the temperature and salinity differences as
\begin{equation}
  \Delta_T = T_{\rm bot} - T_{\rm top}, \quad \quad \Delta_S = S_{\rm top} - S_{\rm bot}.
\end{equation}
The subscripts ``bot'' and ``top'' denote the values at the bottom and top plates, respectively. In this study we always set $\Delta_T<0$ and $\Delta_S>0$, which means that the flow is driven by the salinity difference and stabilized by the temperature difference. The flow control parameters are the Prandtl and Rayleigh numbers
\begin{equation}
  Pr_\zeta = \frac{\nu}{\kappa_\zeta}, \quad \quad 
  Ra_\zeta = \frac{g \beta_\zeta \Delta_\zeta L^3}{\kappa_\zeta \nu},
\end{equation}
with $\zeta=T,S$, or alternatively one can also use the Lewis number and the density ratio
\begin{equation}
  Le = \frac{\kappa_T}{\kappa_S} = \frac{Pr_S}{Pr_T}, \quad\quad
  \Lambda = \frac{\beta_T|\Delta_T|}{\beta_S|\Delta_S|} = \frac{Le |Ra_T|}{|Ra_S|}.
\end{equation}
Note that $\Lambda$ measures the ratio between the buoyancy force induced by the two scalars. $\Lambda=0$ corresponds to a Rayleigh-B\'{e}nard flow purely driven by salinity difference. As $\Lambda$ increases, the magnitude of the stabilizing buoyancy force induced by temperature difference becomes stronger when compared to the destabilising force of the salinity difference. For otherwise fixed parameters, one can view $\Lambda$ also as dimensionless (stabilising) temperature difference between top and bottom plates.

In all our simulations the Prandtl numbers are fixed at $Pr_T=7$ and $Pr_S=700$, i.e. the typical values of seawater. The Lewis number is then $Le=100$. The density ratio is in the range of $0.1\le\Lambda\le10$, which falls into the salt-finger regime. Initially the fluid is at rest. The temperature has a vertically linear distribution, and salinity is uniform and equal to $(S_{\rm bot}+S_{\rm top})/2$, respectively. These initial fields are similar to those in the experiments of~\cite{Hage_Tilgner2010} and~\cite{Kellner_Tilgner2014}. Small random perturbations are superposed to both scalar fields to accelerate the development of the flow. The parameters explored in the present work are shown in figure~\ref{fig:param} and more details can be found in the Appendix. The numerical method is reported in~\cite{multigrid2015}. The typical flow structures are similar to those shown in our previous study (e.g.~see figure~1 of~\cite{ddcjfm2015}). The bulk region is dominated by salt fingers, and thin boundary layers develop for the velocity and salinity fields adjacent to two plates.
\begin{figure}
\centering
\includegraphics[width=\textwidth]{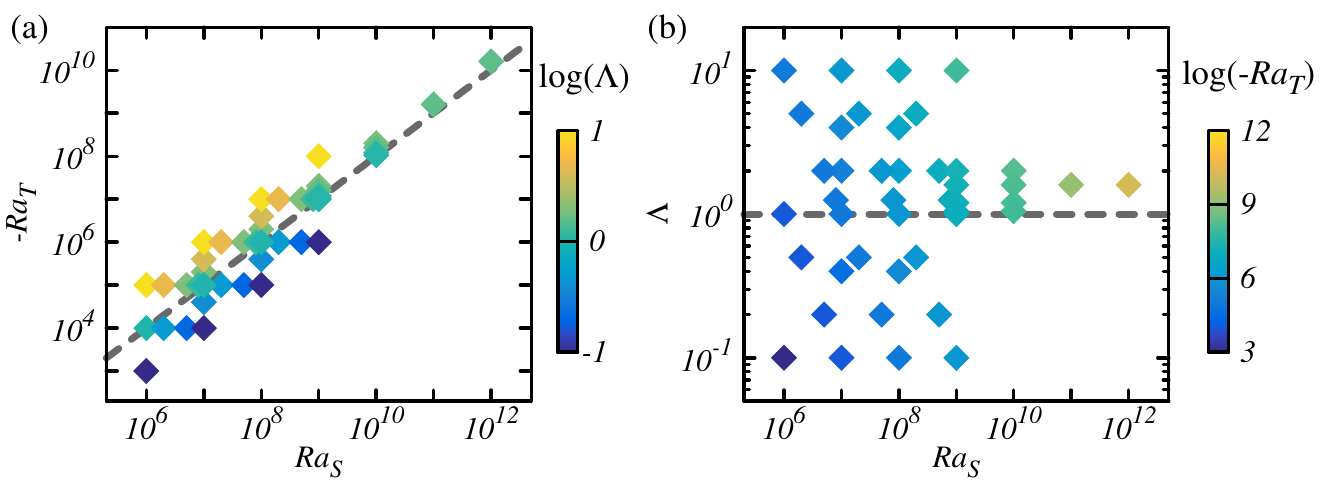}%
\caption{Explored parameters shown (a) on the $Ra_S$--$Ra_T$ plane and coloured by $\log(\Lambda)$, and (b) on the $Ra_S$--$\Lambda$ plane and coloured by $\log(-Ra_T)$. The dashed lines indicate $\Lambda=1$, i.e. the fluid layer is neutrally buoyant. The fluid is overall unstable for the parameters below the lines and stable for those above the lines, respectively.}
\label{fig:param}
\end{figure}

\section{Global responses and the GL theory}

\begin{figure}
\centering
\includegraphics[width=\textwidth]{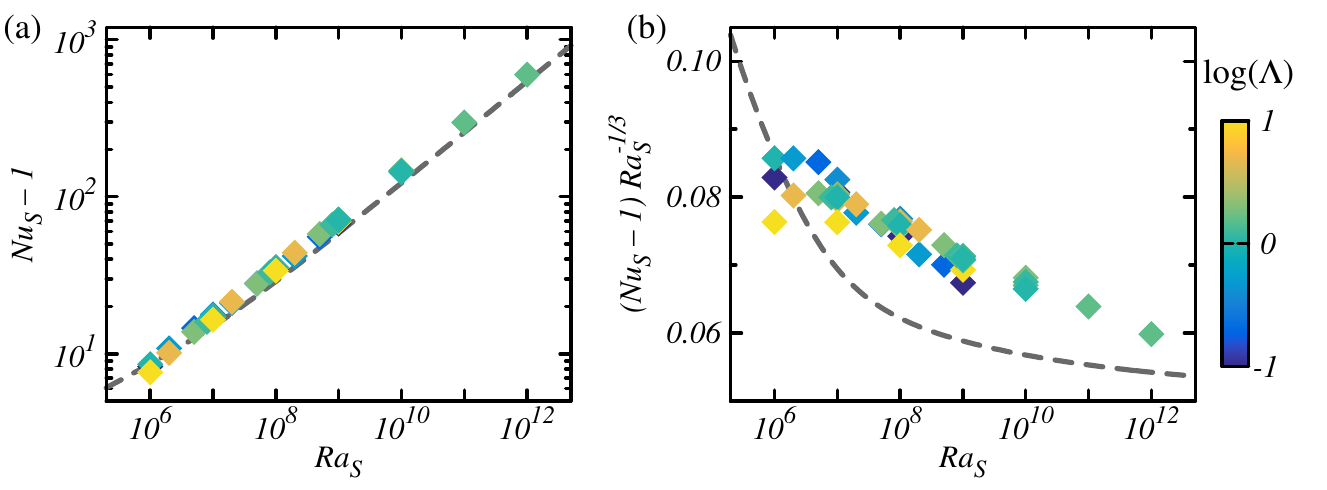}%
\caption{Dependences of the convective salinity flux $\Nus_S-1$ on $\Ray_S$ in (a) a log-log plot and (b) compensated form. Symbols are coloured according to the logarithm of $\Lambda$. The dashed lines are the predictions of the GL theory.}
\label{fig:rasnus}
\end{figure}

The three key responses of the system are the salinity and heat fluxes and the flow velocity, which are measured in non-dimensional form by the Nusselt and Reynolds numbers as
\begin{equation}\label{eq:Nuss}
  \Nus_S = \frac{\langle u_3 s \rangle - \kappa_S \partial_3\langle s \rangle}{\kappa_S \Delta_S L^{-1}},
 \quad\quad
  \Nus_T = \frac{\langle u_3 \theta \rangle - \kappa_T \partial_3\langle \theta \rangle}{\kappa_T \Delta_T L^{-1}},
 \quad\quad 
  \Rey = \frac{u_{rms} L}{\nu}.
\end{equation}
Here $\langle\cdot\rangle$ stands for the average over the entire domain and time. $u_{rms}$ denotes the rms value of the velocity magnitude. In figure~\ref{fig:rasnus} we plot $\Nus_S$ versus $\Ray_S$, both as a log-log plot and in compensated form. We compare the numerical results to the GL theory with the updated coefficients~\citep{GLrefit2013}. Note that there is no new free parameter here and the curve is fully determined by the theory. In our previous study~\citep{ddcjfm2015} we have shown, for a smaller parameter range, that $\Nus_S$ mainly depends on $\Ray_S$ and it shows only a minor dependence on $\Lambda$. The variation of $\Nus_S$ with $\Ray_S$ is also close to the GL prediction. Indeed, for the current dataset with a much larger parameter range, $\Nus_S$ still follows a single trend which is close to the GL prediction. The small discrepancy between the numerical results and the GL theory was also found for standard RB flow when one applies the GL theory at high Prandtl number, e.g. see figure 7 of~\cite{GLrefit2013}. 

\begin{figure}
\centering
\includegraphics[width=\textwidth]{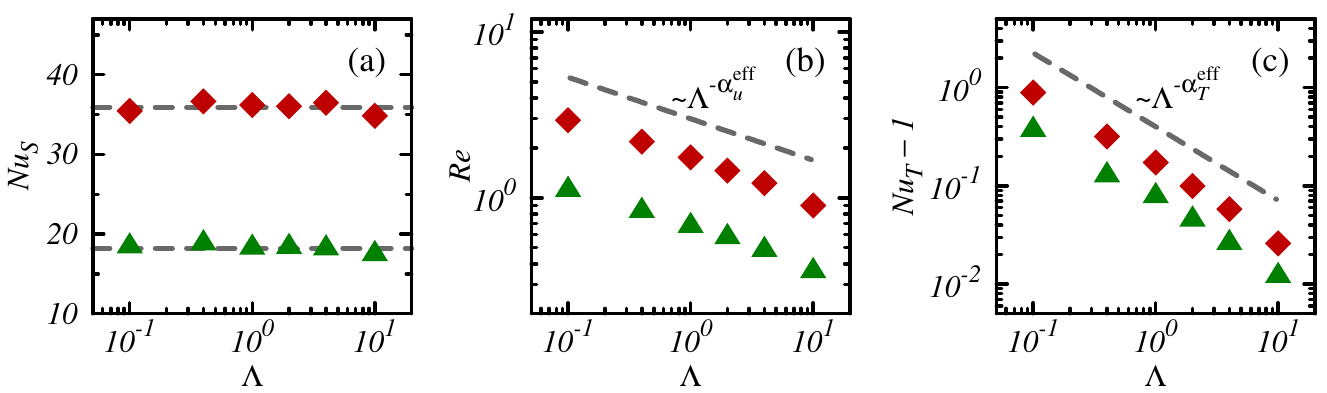}%
\caption{Dependences of (a) $\Nus_S$ and (b) $\Rey$ on $\Lambda$ for two fixed $\Ray_S=10^7$ (green triangles) and $10^8$ (red diamonds). In (a) the two horizontal dashed lines mark $\Nus_S=18.2$ and $35.9$, which are the averaged values over different $\Lambda$'s for each $\Ray_S$, respectively. In (b) the dashed line indicates the scaling relation $\Lambda^{-\alpha^{\rm eff}_u}$ with $\alpha^{\rm eff}_u=0.25$. In (c) the dashed line indicates the scaling relation $\Lambda^{-\alpha^{\rm eff}_T}$ with $\alpha^{\rm eff}_T=0.75$.}
\label{fig:dreffs}
\end{figure}

Unlike on the salinity flux, however, the density ratio $\Lambda$ has strong influence on the flow velocity and  on the heat transfer. To reveal the effects of $\Lambda$ on the system responses, we focus on two sets of cases with $\Ray_S$ fixed at $10^7$ and $10^8$, respectively. Each set consists of six cases with $\Lambda$ ranging from $0.1$ to $10$. In figure~\ref{fig:dreffs} we plot the dependences of $\Nus_S$, $\Rey$, and $\Nus_T$ on $\Lambda$ for these two sets. Clearly, when $\Ray_S$ is fixed, $\Nus_S$ is almost constant for the range of $\Lambda$ considered here. In contrast, for fixed $\Ray_S$ both the Reynolds number $\Rey$ and the convective heat transfer $(\Nus_T-1)$ decrease as $\Lambda$ increases from $0.1$ to $10$. The different behaviors of $\Nus_S$ and $\Rey$ versus $\Lambda$ suggest that for fixed salinity difference, increasing the relative strength of the stabilizing temperature difference suppresses the flow motions. Meanwhile, the flow patterns adjust themselves such that the salinity transfer only changes slightly.

\begin{figure}
\centering
\includegraphics[width=\textwidth]{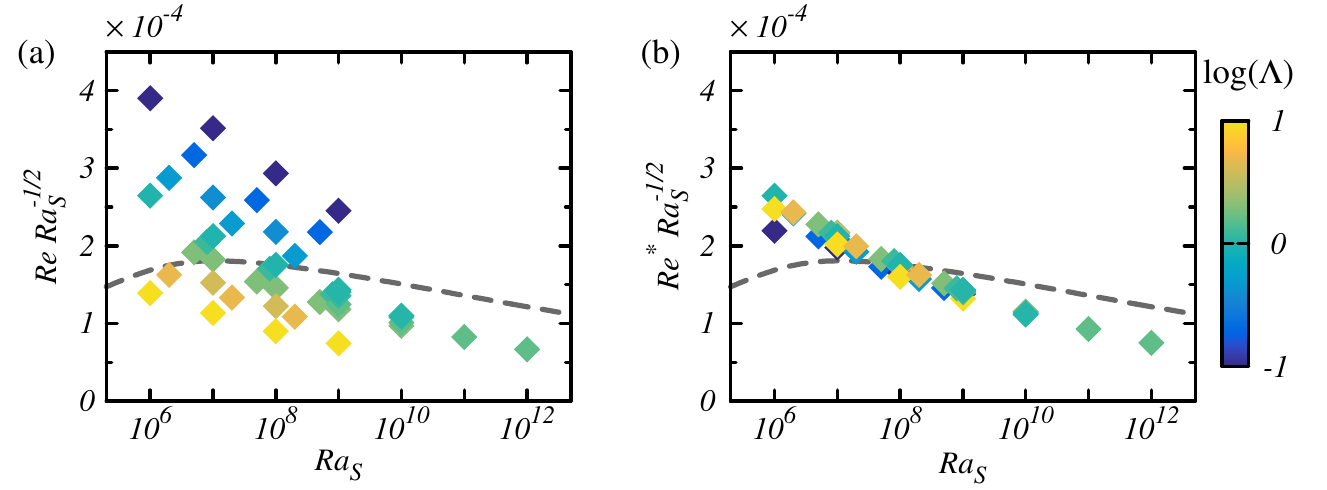}%
\caption{(a) Dependence of $\Rey$ versus $\Ray_S$ for all data points. (b) Dependence of the rescaled Reynolds number $\Rey^*=\Rey\Lambda^{\alpha^{\rm eff}_u}$ with $\alpha^{\rm eff}_u=0.25$ versus $\Ray_S$ for all data points. Symbols are coloured according to the logarithm of $\Lambda$. The dashed lines are the predictions of the GL theory.}
\label{fig:rasrey}
\end{figure}

In figure~\ref{fig:dreffs}(b) we see the Reynolds number $\Rey$ as function of the density ratio $\Lambda$. Note that even for the larger salinity Rayleigh number $\Ray_S=10^8$ and for small $\Lambda$ it is never larger than $3$, far away from any turbulent behavior, but reflecting the laminarity of the flow. Also the thermal flux is very small, with $\Nus_T-1$ always being smaller than $1$, close to the value for pure molecular diffusion, i.e.~$\Nus_T=1$. From figures~\ref{fig:dreffs}(b,\,c) one also observes that the dependences of $\Rey$ and $(\Nus_T-1)$ on $\Lambda$ follow certain power-law scalings. Thus a possible strategy for constructing scaling laws for those two quantities would be first extracting the scaling laws of the $\Lambda$-dependences of $\Rey$ and $(\Nus_T-1)$ for fixed $\Ray_S$, and then examining their behaviours for different $\Ray_S$. This strategy is applied to $\Rey$ and $(\Nus_T-1)$ as follows. Let's assume that $\Rey$ scales as $\Lambda^{-\alpha^{\rm eff}_u}$ for fixed $\Ray_S$, where $\alpha^{\rm eff}_u$ is a positive exponent. The minus sign is introduced since $\Rey$ decreases as $\Lambda$ increases. The value of $\alpha^{\rm eff}_u$ can be determined by the data points in figure~\ref{fig:dreffs}(b). For the set of $\Ray_S=10^7$, the exponent given by a linear regression is $\alpha^{\rm eff}_u=0.24\pm0.02$. And for the set of $\Ray_S=10^8$ it is $\alpha^{\rm eff}_u=0.25\pm0.02$. The two values are very close to each other. To obtain a single value for $\alpha^{\rm eff}_u$, we first shift the data points vertically in figure~\ref{fig:dreffs}(b) such that the two points at $(\Lambda,~\Ray_S)=(1,10^7)$ and $(1,10^8)$ collapse and then conduct a linear regression with all $12$ data points. The final value is $\alpha^{\rm eff}_u=0.25\pm0.02$. In figure~\ref{fig:rasrey}(a) we plot the original values of $\Rey$ compensated by $\Ray_S^{-1/2}$. The data points are scattered since $\Rey$ is strongly affected by $\Lambda$. However, we can define a rescaled Reynolds number $\Rey^*=\Rey\Lambda^{\alpha^{\rm eff}_u}$ with $\alpha^{\rm eff}_u=0.25$. In figure~\ref{fig:rasrey}(b) we plot $\Rey^*$ against $\Ray_S$ for the whole dataset. All the data points collapse and the dependence $\Rey^*(\Ray_S)$ is quite close to the GL prediction. 

\begin{figure}
\centering
\includegraphics[width=\textwidth]{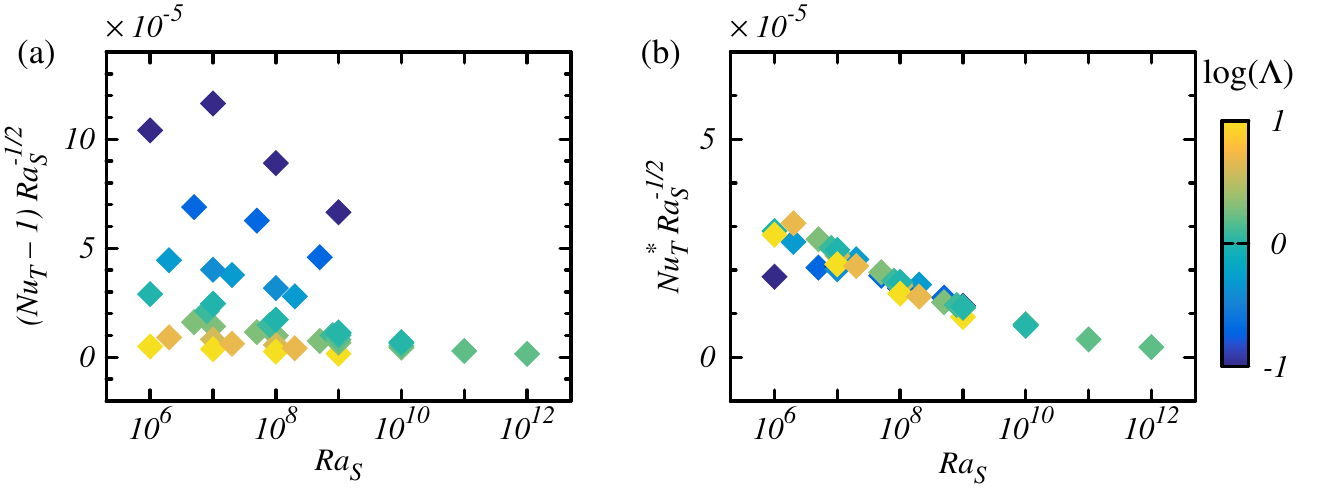}%
\caption{(a) Dependence of the convective heat flux $\Nus_T-1$ versus $\Ray_S$. (b) Dependence of the rescaled heat Nusselt number $\Nus_T^*=(\Nus_T-1)\Lambda^{\alpha^{\rm eff}_T}$ with $\alpha^{\rm eff}_T=0.75$ versus $\Ray_S$. Symbols are coloured according to the logarithm of $\Lambda$. Note that the range of the vertical axis in (b) is half of that in (a).}
\label{fig:rasnut}
\end{figure}

A similar analysis can be done for $\Nus_T$. We assume that the convective heat flux scales as $(\Nus_T-1) \sim \Lambda^{-\alpha^{\rm eff}_T}$ and calculate the value of $\alpha^{\rm eff}_T$ from figure~\ref{fig:dreffs}(c) by following the same procedure for $\alpha^{\rm eff}_u$. The final value is $\alpha^{\rm eff}_T=0.75\pm0.03$. And a rescaled convective heat Nusselt number can be defined as $\Nus_T^*=(\Nus_T-1)\Lambda^{\alpha^{\rm eff}_T}$ with $\alpha^{\rm eff}_T=0.75$. In figure~\ref{fig:rasnut} we plot the original value of $\Nus_T-1$ and the rescaled ones $\Nus_T^*$ against $\Ray_S$ in a compensated form. Again, the original values are scattered, but the rescaled quantity $\Nus_T^*$ follows a single trend versus $\Ray_S$. Two things should be noted from figure~\ref{fig:rasnut}. First, the quantities are compensated by $\Ray_S^{-1/2}$ as for $\Rey$, instead of $\Ray_S^{-1/3}$ as for $\Nus_S$. Second, the dependences of $\Nus_T^*$ on $\Ray_S$ shown in figure~\ref{fig:rasnut}(b) is very similar to that of $\Rey^*$ shown in figure~\ref{fig:rasrey}(b). A reasonable argument is that in the current flow setup, the temperature field is stabilizing rather than driving the flow. Thus the convective heat flux is generated as the temperature anomaly is ``passively'' carried by the flow motions which are sustained by the salinity difference. Therefore $\Nus_S$ behaves similarly to that of a RB flow, while $\Nus_T^*$ exhibits similar scaling as $\Rey^*$.

\begin{figure}
\centering
\includegraphics[width=\textwidth]{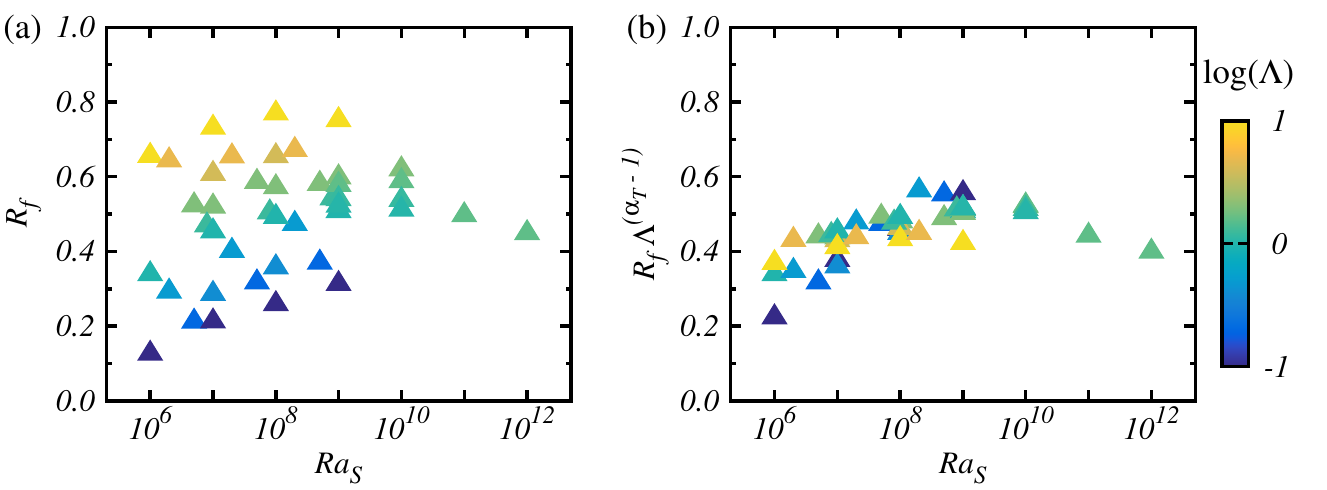}%
\caption{(a) Dependence of the density flux ratio $R_f$ versus $\Ray_S$. (b) Dependence of the rescaled density flux ratio $R_f^*=R_f \Lambda^{\alpha^{\rm eff}_T-1}$ with $\alpha^{\rm eff}_T=0.75$ versus $\Ray_S$. Symbols are coloured according to the logarithm of $\Lambda$.}
\label{fig:rasdfr}
\end{figure}

Although the non-dimensional thermal flux is much smaller than the non-dimensional salinity flux, the density-anomaly flux associated with the temperature field may not be negligible when compared to that associated with the salinity field, because of the huge difference between the two molecular diffusivities. This can be seen from the density flux ratio which defined as
\begin{equation}\label{eq:Rflx}
  R_f = \frac{\beta_T\langle u_3 \theta \rangle_V}{\beta_S\langle u_3 s \rangle_V}
      = \Lew\,\Lambda\,\frac{\Nus_T-1}{\Nus_S-1}.
\end{equation}
In figure~\ref{fig:rasdfr}(a) we plot the variation of $R_f$ versus $\Ray_S$. The density flux ratio $R_f$ has the value between $0.1$ and $0.8$. As the density ratio $\Lambda$ increases (symbol color changes from dark blue to light yellow), $R_f$ also increases. In previous discussion we showed that, for fixed $\Ray_S$, $\Nus_T-1$ scales as $\Lambda^{-\alpha^{\rm eff}_T}$ and $\Nus_S$ is nearly constant. Then by definition~(\ref{eq:Rflx}) one expects $R_f\sim\Lambda^{1-\alpha^{\rm eff}_T}$ for fixed $\Ray_S$. In figure~\ref{fig:rasdfr}(b) we plot the rescaled density flux ratio $R^*_f=R_f\Lambda^{\alpha^{\rm eff}_T-1}=R_f\Lambda^{-0.25}$, and indeed all the data points collapse. 

\section{Flow structures and thicknesses of boundary layers and fingers} 

\begin{figure}
\centering
\includegraphics[width=\textwidth]{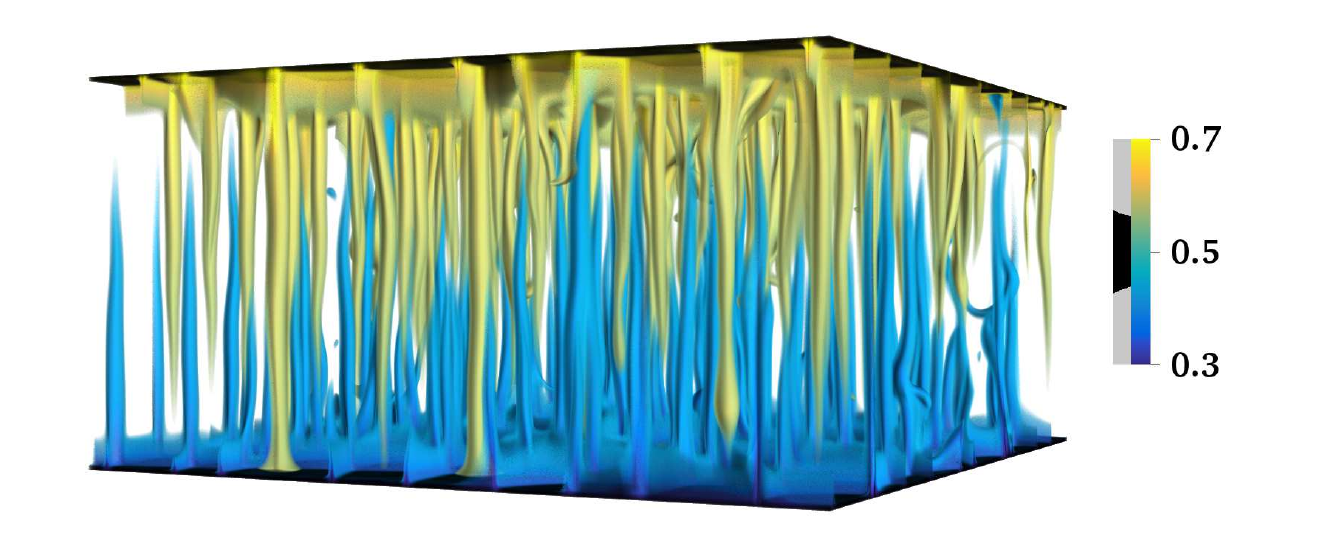}%
\caption{Typical flow structures shown by three-dimensional volume rendering of the salinity field with the flow control parameters $\Ray_T=10^6$ and $\Ray_S=10^8$, or equivalently $\Lambda=1$. The color and opacity are both determined by salinity.}%
\label{fig:fg3d}
\end{figure}

We now turn to the flow structures. We will focus on the horizontal width of salt fingers and the thicknesses of velocity and salinity boundary layers. Then we will show the horizontal zonal flows which were observed at very high Rayleigh numbers. In all our simulations salt fingers develop in the bulk of the flow domain. A typical flow field can be seen in figure~\ref{fig:fg3d}, which shows the three-dimensional volume rendering of the salinity field at $\Ray_T=10^6$ and $\Ray_S=10^8$ (equivalently $\Lambda=1$). The finger layer is bounded by two thin boundary layers adjacent to both plates. The individual salt fingers can be distinguished. 

\subsection{Salinity boundary layers}

\begin{figure}
\centering
\includegraphics[width=0.7\textwidth]{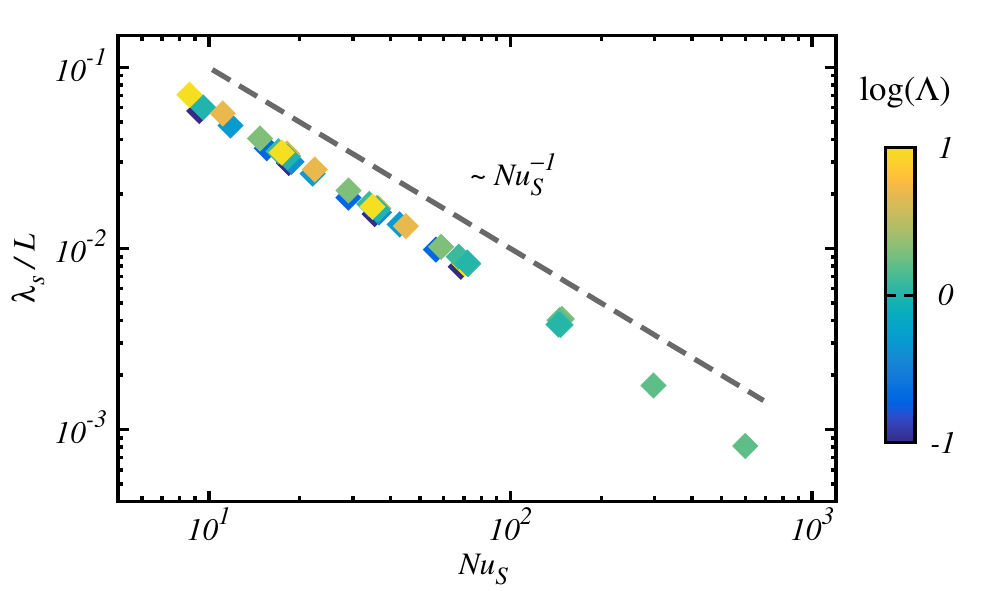}%
\caption{The thickness of the salinity boundary $\lambda_s/L$ versus the salinity Nusselt number $\Nus_S$. The dashed line has the slope $-1$. Symbols are coloured according to the logarithm of $\Lambda$.}
\label{fig:nussbl}
\end{figure}

The thickness of the salinity boundary layer $\lambda_s$ is defined as the distance from the plate to the location of the first peak of $\sigma_s(z)$, which is the standard deviation of salinity over every horizontal plane. In figure~\ref{fig:nussbl} we plot the dependence of $\lambda_s/L$ on $\Nus_S$, and find that it perfectly scales as $\Nus_S^{-1}$. The scaling $\lambda_s/L \sim \Nus_S^{-1}$ can be understood by following the argument for RB flows at high Prandtl numbers~\citep{GL2001,GL2002}. The salinity Nusselt number is related to the salinity dissipation rate by the exact relation 
\begin{equation}
   \epsilon_s \equiv \left\langle \kappa_S [\partial_i s]^2 \right\rangle_V 
                  = \kappa_S\, (\Delta_S)^2\, L^{-2}\, \Nus_S.
\end{equation}
For the salinity field with high Prandtl number, the dissipation is dominated by the contribution from the boundary layers. The volume integral of $\epsilon_s$ in the two boundary layers may be approximated as
\begin{equation}
  \epsilon_s \sim \epsilon^{\rm BL}_s 
    = \left\langle \kappa_S [\partial_i s]^2 \right\rangle_{\rm BL} 
    \sim \kappa_S \left(\frac{\Delta_S}{2\lambda_s}\right)^2 \frac{2\lambda_s}{L} 
    = \kappa_S \frac{\Delta^2_S}{2\lambda_s L}.
\end{equation}
Combining the above two equations, one readily obtains $\lambda_s/L \sim \Nus_S^{-1}$, which is exactly the case as shown in figure~\ref{fig:nussbl}.

\subsection{The width of the salt fingers}

\begin{figure}
\centering
\includegraphics[width=0.8\textwidth]{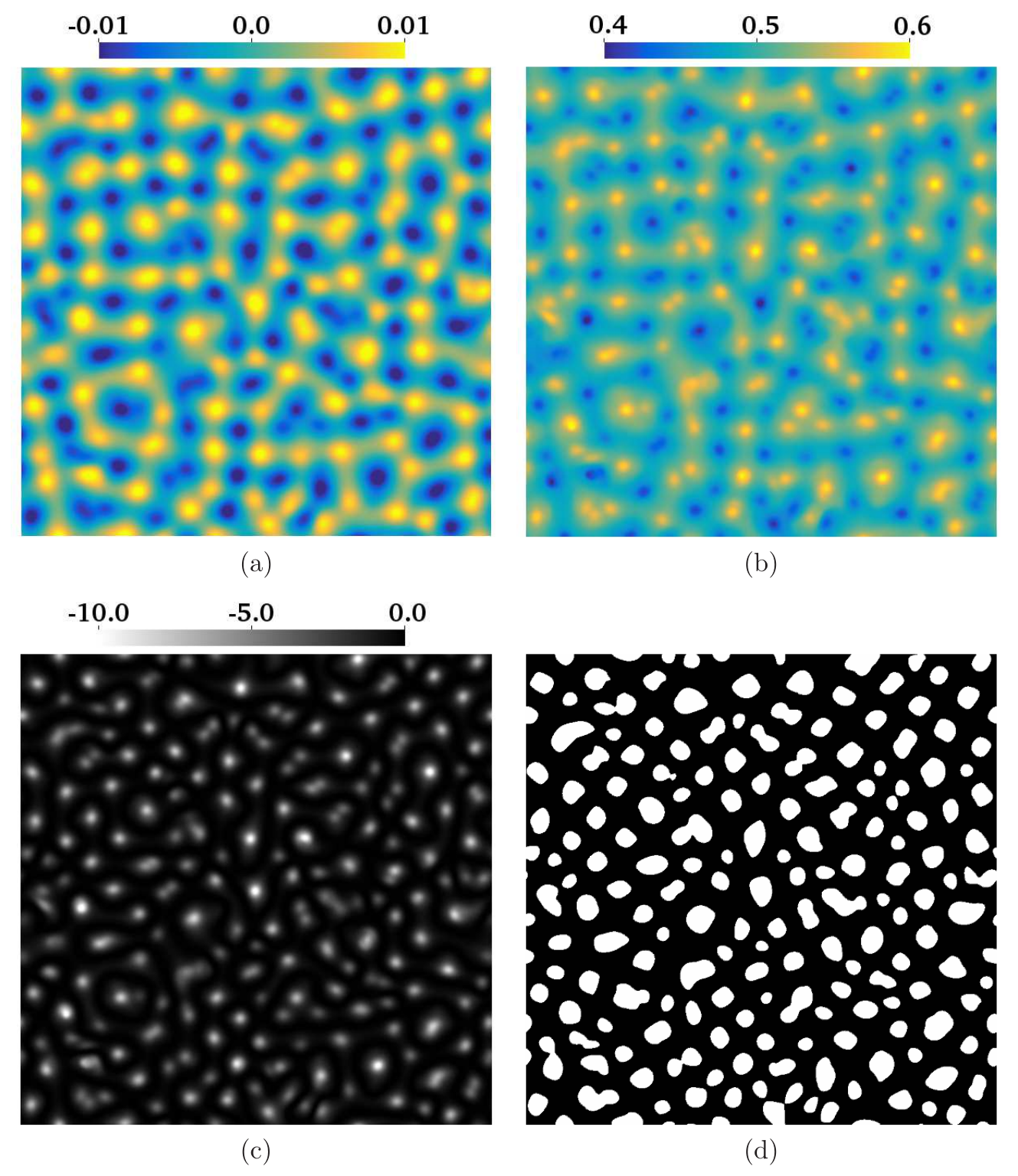}%
\caption{An example of the raw data on the mid plane $z/L=0.5$ and corresponding finger detection. The control parameters are $\Ray_T=10^6$ and $\Ray_S=10^8$, or equivalently $\Lambda=1$. For this flow the standard deviations for the vertical velocity $u_3$ is $\sigma_{u_3}=4.9\times10^{-3}$ and for the salinity $s$ is $\sigma_s=2.9\times10^{-2}$, respectively. (a,b) Contours of $u_3$ and $s$, (c) the cross correlation coefficient between the variances of $u_3$ and $s$, i.e. $(u'_3 s')/(\sigma_{u_3}\sigma_s)$,  and (d) the finger regions with $|s'|>\sigma_s$ and $|u'_3|>\sigma_{u_3}$, as shown by the white patches.}%
\label{fig:fgreg}
\end{figure}

The horizontal length scale of the salt fingers can be determined from the flow field on the horizontal mid-plane $z/L=0.5$. In figures~\ref{fig:fgreg}(a,b) we show the contours of the vertical velocity $u_3$ and the salinity $s$ on the horizontal plane $z/L=0.5$ for the same flow field as shown in figure~\ref{fig:fg3d}. Clearly, most fingers have almost circular shape in the horizontal sections. Some sheet-like links can be found, connecting different fingers, but usually they are not pronounced. 

Large salinity anomaly is usually transported by salt fingers, which can be clearly seen from the cross correlation between the variance of the vertically velocity $u'_3$ and that of the salinity $s'$. The prime denotes the deviation of the quantity from its mean value over the plane. In figure ~\ref{fig:fgreg}(c) we plot the cross correlation coefficient function $C=(u'_3 s')/(\sigma_{u_3}\sigma_s)$ with $\sigma$ denoting the standard deviation of the quantity. $C$ has large negative values at isolated regions, because the ascending (descending) fingers carry negative (positive) salinity anomaly. The global correlation coefficient, which is the average of $C$, is $-0.98$. Thus $u'_3$ and $s$ are almost perfectly negatively correlated, namely, salt fingers dominate the salinity transfer in the bulk region. From the flow fields as shown in figures~\ref{fig:fgreg}(a,b), we can identify the salt-finger regions, which are the regions with $|s'|>\sigma_s$ and $|u'_3|>\sigma_{u_3}$. In figure~\ref{fig:fgreg}(d) we present the finger regions detected by this method. Each white patch corresponds to an individual finger region. 

Such flow fields were stored constantly during each simulation, and the horizontal length scale of the salt finger $d$ is extracted from those data. First, the total number of the finger regions $N_f$ is counted and averaged over time, and the average spatial area occupied by one finger is calculated as $A_f = A_{\rm total} / N_f$ with $A_{\rm total}$ being the total horizontal area of the domain. The finger width $d$ is then calculated from $A_f$ by assuming that the finger has circular shape, i.e. $d=\sqrt{4A_f/\pi}$. Thus $d$ is actually the average diameter of the slender convection cells associated with salt fingers.  

\begin{figure}
\centering
\includegraphics[width=0.6\textwidth]{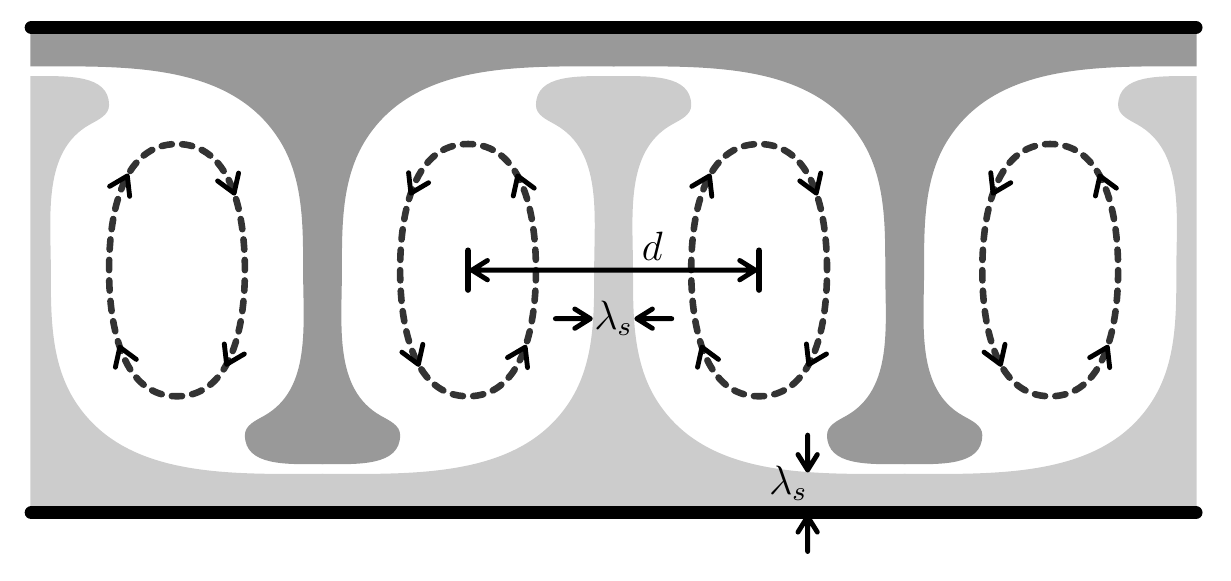}%
\caption{A sketch to demonstrate the mean salinity within finger-convection cells.}%
\label{fig:sketch}
\end{figure}

\begin{figure}
\centering
\includegraphics[width=0.6\textwidth]{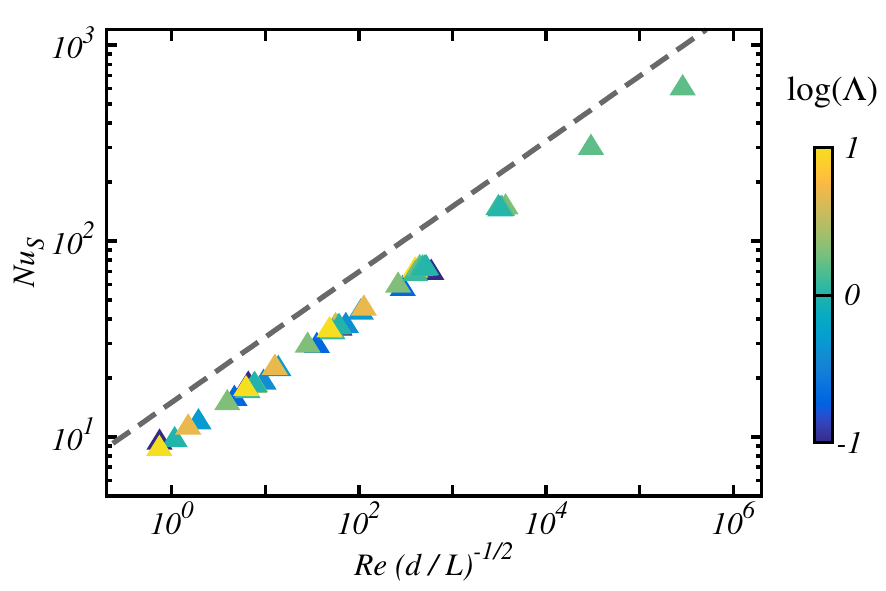}%
\caption{The scaling between $\Rey (d/L)^{-2}$ and $\Nus_S$. Symbols are coloured according to the logarithm of $\Lambda$. The dashed line has a slope of $1/3$.}%
\label{fig:fgwnus}
\end{figure}

Following the argument in~\citet{Hage_Tilgner2010}, we can relate the scaling behaviour of the finger width to that of the Reynolds number. Since the salinity is mainly transferred by fingers, the salinity Nusselt number can be approximated as
\begin{equation}\label{eq:nusf}
  \Nus_S \approx \frac{U^f \, \overline{s^f}}{\kappa_S \Delta_S L^{-1}},
\end{equation}
in which $U^f$ is the characteristic velocity of the fingers and $\overline{s^f}$ is the mean salinity anomaly within the fingers, respectively. The salinity anomaly is carried from the boundary layers and the side diffusion is weak in the bulk due to the large Prandtl number. Then within each finger convection cell of width $d$, the salinity core has similar width as the thickness of salinity boundary layer, i.e.~$\lambda_s$, as shown by the sketch in figure~\ref{fig:sketch}.
The mean salinity anomaly inside a finger can then be calculated, for circular fingers, as~\citep{Hage_Tilgner2010}
\begin{equation}\label{eq:sfa}
   \overline{s^f} = \frac{\Delta_S}{2} \left( \frac{\lambda_s}{d} \right)^2.
\end{equation}
Then combining the above two equations, and considering that $\lambda_s / L \sim \Nus_S^{-1}$ as proven in the previous section, one obtains
\begin{equation}\label{eq:sfa}
    \frac{U^f L}{\kappa_S} \frac{L^2}{d^2} \sim \Nus_S^3,  \quad \mbox{or} \quad 
    \Rey \left(\frac{d}{L}\right)^{-2} \sim \Nus_S^3.
\end{equation}
This scaling relation can be clearly seen in figure~\ref{fig:fgwnus}, in which we plot $\Rey (d/L)^{-2}$ versus $\Nus_S$ for all cases.

\begin{figure}
\centering
\includegraphics[width=0.6\textwidth]{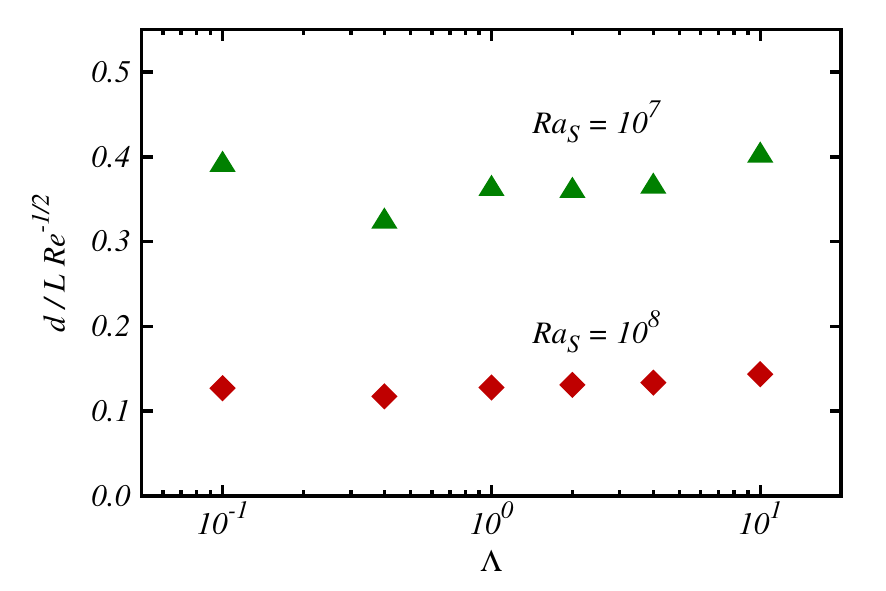}%
\caption{The dependence of the finger width $d/L$, rescaled as $(d/L)\Rey^{-1/2}$, on $\Lambda$ for two fixed $\Ray_S=10^7$ and $10^8$.}%
\label{fig:drfgw}
\end{figure}

\begin{figure}
\centering
\includegraphics[width=\textwidth]{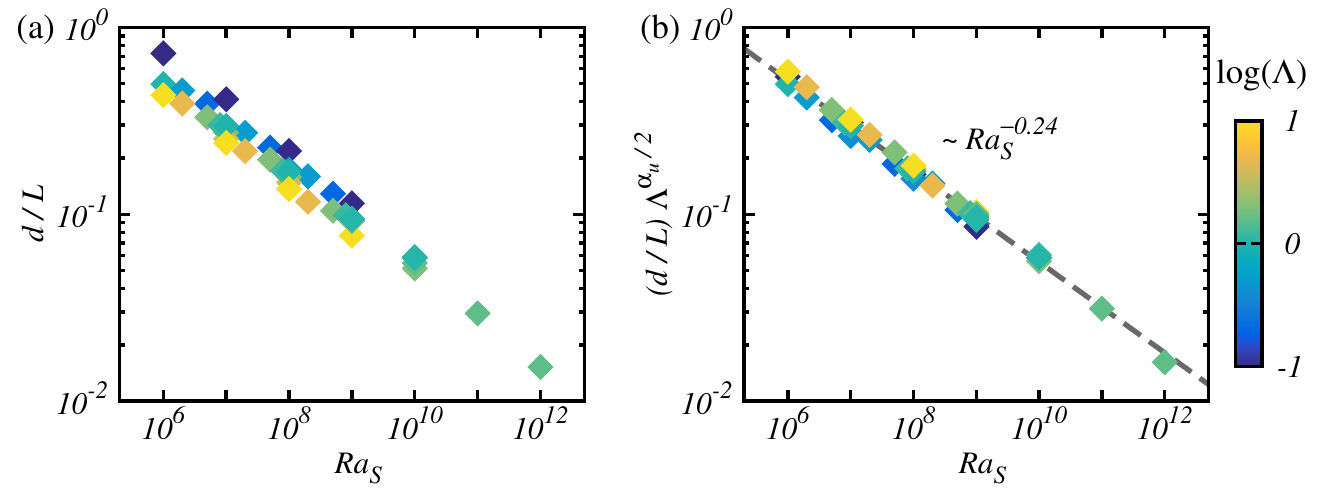}%
\caption{The dependences of (a) the original finger width $d/L$ and (b) the rescaled finger width $(d/L)\Lambda^{\alpha^{\rm eff}_u/2}$ with $\alpha^{\rm eff}_u=0.25$ on $\Ray_S$. Symbols are coloured according to the logarithm of $\Lambda$. In (b) the data points collapse after the rescaling, and the dashed line represents the linear regression with a slope of $-0.24$.}%
\label{fig:rasfgw}
\end{figure}

One important consequence of (\ref{eq:sfa}) is, to generate similar salinity flux $\Nus_S$, one must have $d/L\sim\Rey^{1/2}$. For fixed $\Ray_S$ the salinity Nusselt number $\Nus_S$ is almost constant. Then the rescaled finger width $(d/L)\Rey^{-1/2}$ should also be constant. In figure~\ref{fig:drfgw} we plot the rescaled finger width $(d/L)\Rey^{-1/2}$ for two sets of cases with fixed $\Ray_S=10^7$ and $10^8$, respectively. Indeed, when $\Ray_S$ is fixed, $(d/L)\Rey^{-1/2}$ is constant for different $\Lambda$, which confirms the scaling $d/L \sim \Rey^{1/2}$. Furthermore, we have shown in the previous section that $\Rey\sim\Lambda^{-\alpha^{\rm eff}_u}$ with $\alpha^{\rm eff}_u=0.25$ for fixed $\Ray_S$, e.g.~see figure~\ref{fig:dreffs}(b). This implies that $d/L\sim\Lambda^{-\alpha^{\rm eff}_u/2}$ for fixed $\Ray_S$. In figure~\ref{fig:rasfgw} we show the dependences of both the original finger width $d/L$ and the rescaled value $(d/L) \Lambda^{\alpha^{\rm eff}_u/2}$ on $\Ray_S$. The non-dimensional finger width collapses and follows a single power-law scaling when rescaled by $\Lambda^{\alpha^{\rm eff}_u/2}$. A liner regression gives an exponent of $-0.24\pm0.03$, see figure~\ref{fig:rasfgw}(b).

\subsection{Velocity boundary layers}

\begin{figure}
\centering
\includegraphics[width=\textwidth]{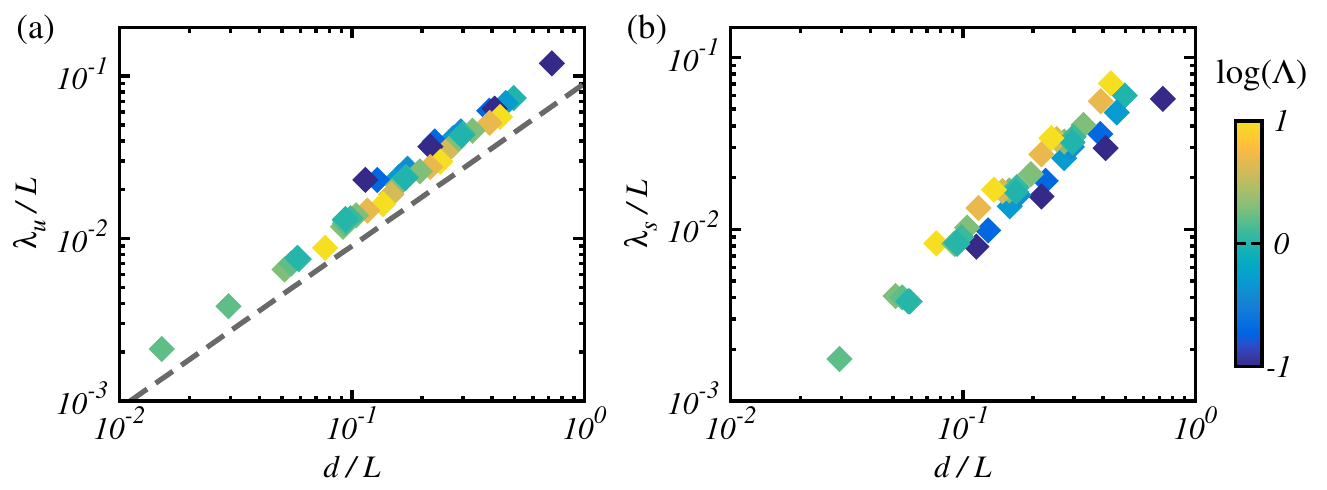}%
\caption{Boundary layer thicknesses of (a) velocity $\lambda_u/L$ and (b) salinity $\lambda_s/L$ versus finger width $d/L$. Symbols are coloured according to the logarithm of $\Lambda$. In (a) the dashed line has a slope unity. }
\label{fig:fwgbls}
\end{figure}

For the current flow, there is no large scale circulation in the bulk. The vertically moving fingers induce converging or diverging flows in the horizontal directions when they move away from or towards the plates. Those horizontal motions form the velocity boundary layer. The thickness of velocity boundary layer $\lambda_u$ can be defined as the distance from the boundary to the location of the first peak of $\sigma_{u_h}(z)$. Here $\sigma_{u_h}(z)$ is the standard deviation of one horizontal velocity component over the plane at the height $z$. Since the velocity boundary layer is driven by the vertical motions of the fingers, it is reasonable to expect that $\lambda_u$ scales as the finger width $d/L$, which is confirmed by figure~\ref{fig:fwgbls}(a). For comparison, we also plot the salinity boundary layer thickness $\lambda_s/L$ versus $d/L$ in figure~\ref{fig:fwgbls}(a). The data points are scattered and no single dependence can be found between $\lambda_s/L$ and $d/L$.

\begin{figure}
\centering
\includegraphics[width=\textwidth]{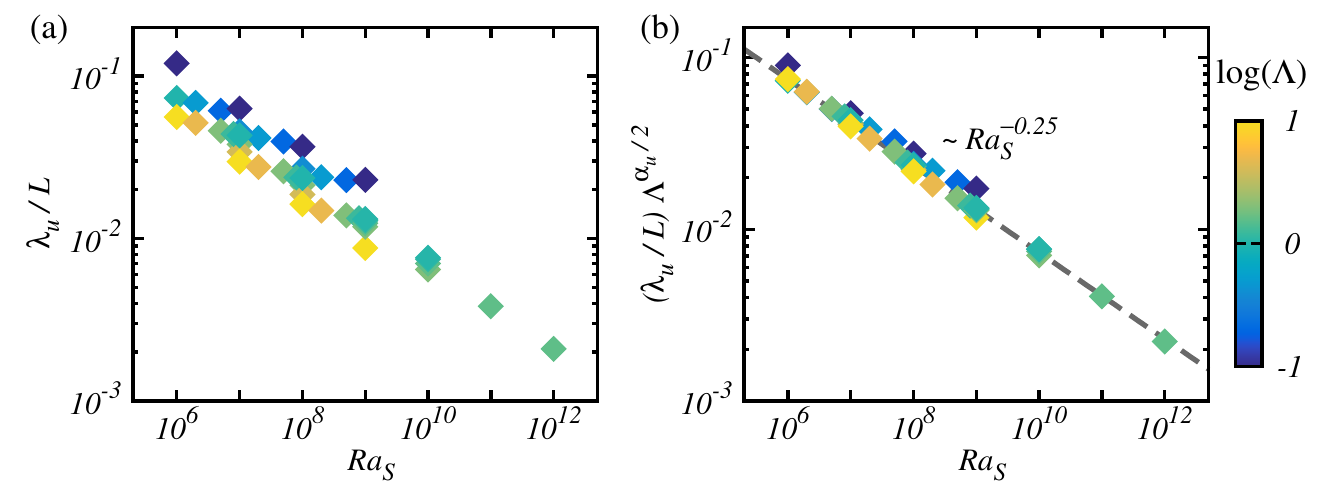}%
\caption{The dependences of (a) the original velocity boundary layer thickness $\lambda_u/L$ and (b) the rescaled velocity boundary layer thickness $(\lambda_u/L)\Lambda^{\alpha^{\rm eff}_u/2}$ with $\alpha^{\rm eff}_u=0.25$ on $\Ray_S$. Symbols are coloured according to the logarithm of $\Lambda$. In (b) the data points collapse after the rescaling, and the dashed line represents the linear regression with a slope of $-0.25$.}
\label{fig:rasubl}
\end{figure}

Moreover, figure~\ref{fig:rasubl} displays the $\Ray_S$-dependence of $\lambda_u/L$, which is affected by both $\Ray_S$ and $\Lambda$, as shown in figure~\ref{fig:rasubl}(a). Similar to the finger width $d/L$, if we rescale $\lambda_u/L$ by $\Lambda^{\alpha^{\rm eff}_u/2}$ with $\alpha^{\rm eff}_u=0.25$, all data points collapse onto a single curve, see figure~\ref{fig:rasubl}(b). The exponent calculated by a linear regression is $-0.25\pm0.02$, which is very close to the exponent for the rescaled finger width as shown in figure~\ref{fig:rasfgw}(b).

\subsection{Horizontal zonal flow at high Rayleigh numbers}

\begin{figure}
\centering
\includegraphics[width=0.32\textwidth]{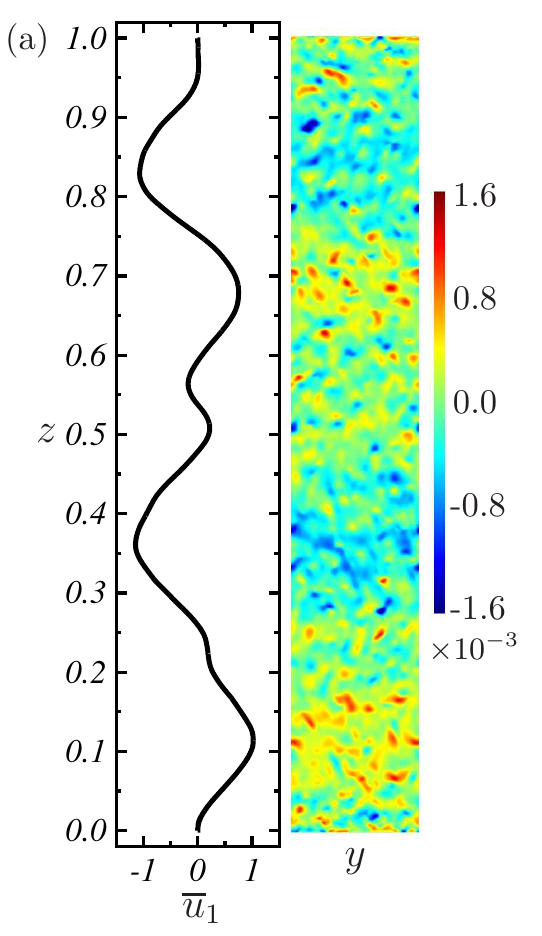}%
~
\includegraphics[width=0.32\textwidth]{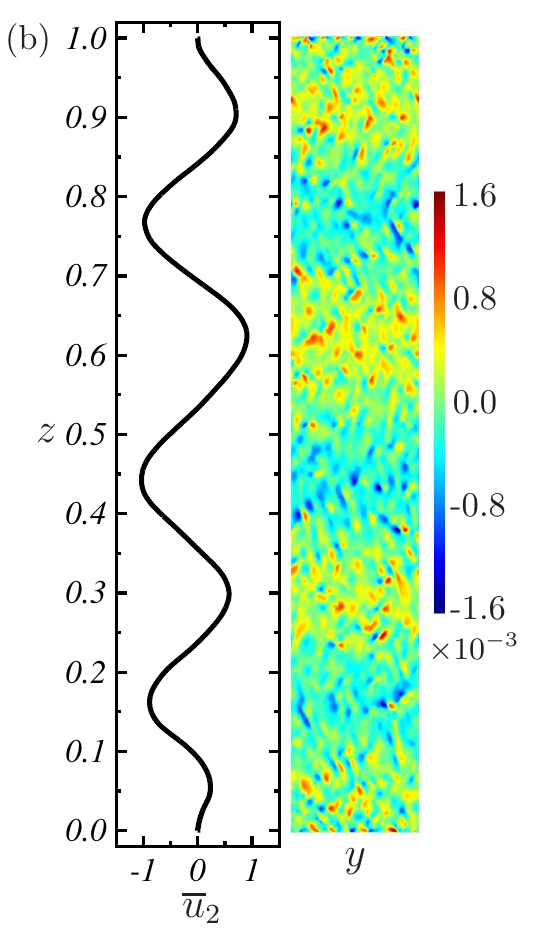}%
\\
\includegraphics[width=0.32\textwidth]{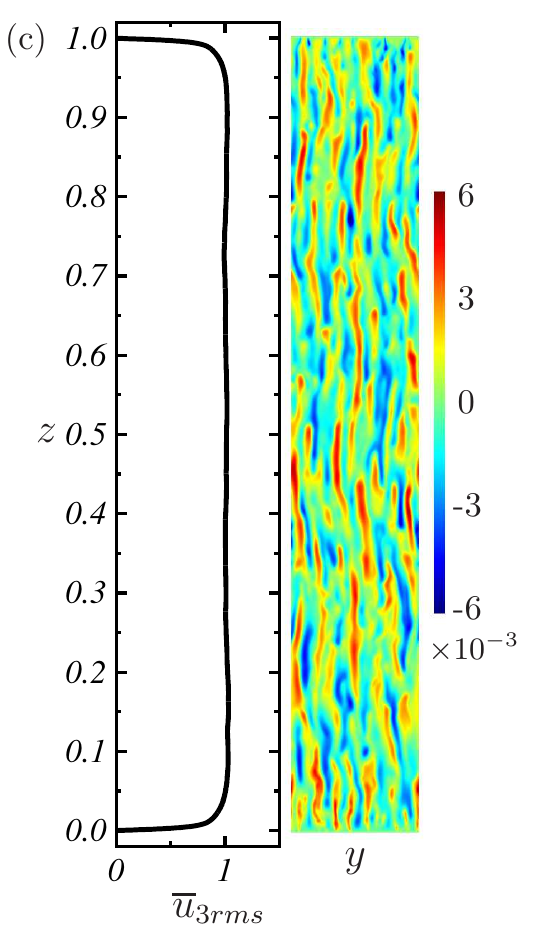}%
~
\includegraphics[width=0.32\textwidth]{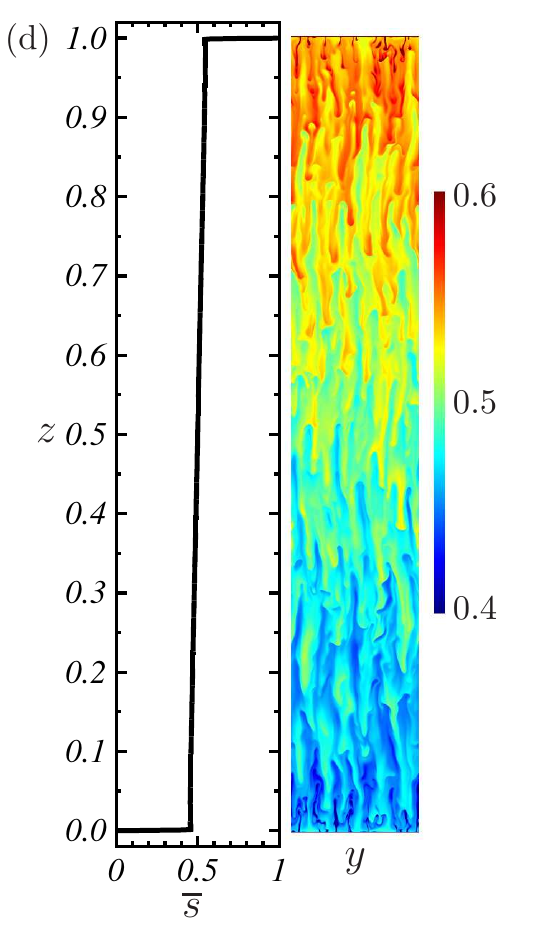}%
\caption{The horizontal zonal flow in the case with $\Ray_S=10^{12}$ and $\Lambda=1.6$. The mean profiles (left) and instantaneous contours on the vertical mid plane (right) are plotted for the two horizontal velocity components $u_1$ (panel a) and $u_2$ (panel b), the vertical velocity component $u_3$ (panel c) and the salinity $s$ (panel d). The mean profiles in (a-c) are normalized by the corresponding volume-averaged rms values $u_{1rms}=3.23\times10^{-4}$, $u_{2rms}=3.18\times10^{-4}$, and $u_{3rms}=1.70\times10^{-3}$, respectively.}%
\label{fig:h32}%
\end{figure}

It is known that large scale oscillating waves can be excited spontaneously in a fully developed salt-finger field and modulate salt fingers, which is attributed to the collective instability~\citep{Stern1969,Holyer1981,Stern2001}. These structures, or the internal gravity waves, were observed in simulations of unbounded domains, such as those reported in~\cite{Stern2001} and \cite{Stellmach2011}. In the fully periodic domain, the phase planes of gravity waves are not necessarily perpendicular to the gravity direction, and these inclined waves cause strong fluctuations in $\Nus_S$ as they travel in space~\citep{Stern2001,Stellmach2011}. In the current study, the inclined waves did {\it not} appear, probably due the vertical constrain of the two plates. Instead, we observed a stack of horizontal zonal flows in alternating directions for the two cases with highest Rayleigh numbers, i.e. $\Ray_S=10^{11}$ and $10^{12}$ at a density ratio $\Lambda=1.6$. In figure~\ref{fig:h32} we show such horizontal zonal flows at $\Ray_S=10^{12}$, which can be clearly seen from the contours of $u_1$ and $u_2$ on the $(y, z)$ mid-plane and the mean profiles in the bulk region, see panels a and b.  

\begin{figure}
\centering
\includegraphics[width=\textwidth]{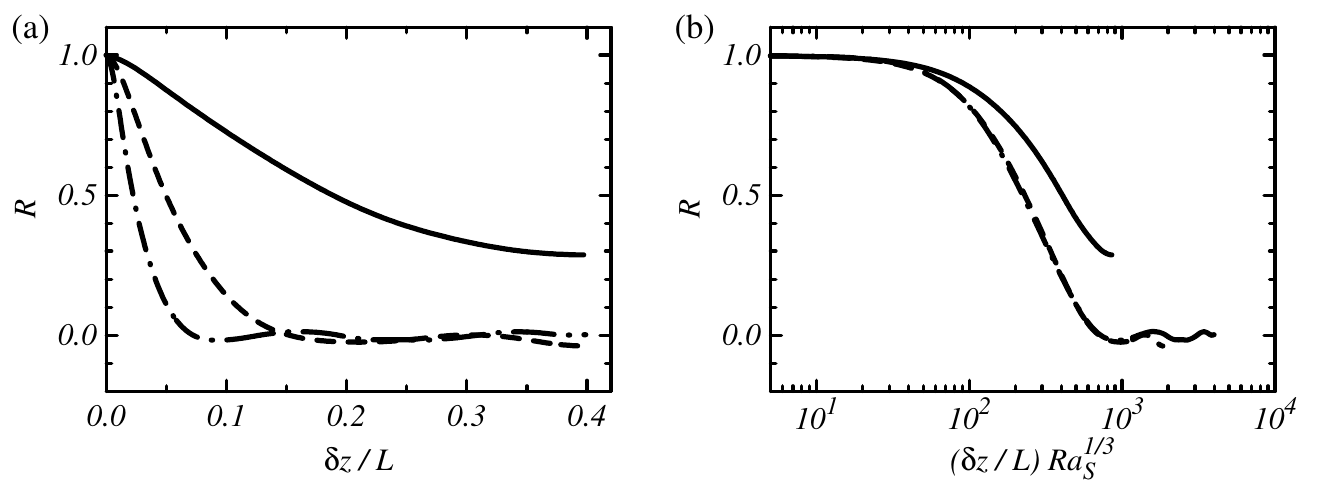}%
\caption{The autocorrelation functions of $u_3$ versus the vertical separations in (a) the global scale $\delta z / L$ and (b) the viscous scale $(\delta z / L)\Ray_S^{1/3}$, respectively. For all three cases the density ratio is fixed to $\Lambda=1.6$. Solid line: $\Ray_S=10^{10}$, dashed line: $\Ray_S=10^{11}$, and dash-dotted line: $\Ray_S=10^{12}$.}
\label{fig:uzcor}
\end{figure}

As $\Ray_S$ becomes large enough, i.e. for tall samples, individual salt fingers do not extend from one plate to the other. To estimate the vertical length of the salt fingers, the autocorrelation function of the vertical velocity $u_3$ is calculated in the vertical direction. We use the flow fields on the $(y,z)$ mid-plane as shown in figure~\ref{fig:h32}. Only the data in the range $0.1<z/L<0.9$ are used to exclude the boundary layer regions near two plates. The autocorrelation functions are defined as
\begin{equation}
  R(\delta z) = \frac{\overline{u_3(y,z,t)u_3(y,z+\delta z,t)}}{\overline{u^2_3(y,z,t)}},
\end{equation}
where the overline denotes the average over $y$, $z$, and $t$ on the $(y,z)$ mid-plane. The autocorrelation $R$ is computed for three cases with $\Lambda=1.6$ and $\Ray_S=10^{10}$, $10^{11}$, and $10^{12}$, respectively. The curves are plotted in figure~\ref{fig:uzcor}(a). For the case with $\Ray_S=10^{10}$, $R$ never decreases to zero, and the fingers can still extend the whole height of the domain. However, for the two cases with larger $\Ray_S$, $R$ does decrease to zero. Thus $u_3$ decorrelates at a height smaller than the domain height, implying that the average height of the salt fingers are smaller than the sample height $L$. From the curves we determine the first zero point of $R$ at $\delta z/L\approx0.15$ for $\Ray_S=10^{11}$, and $\approx0.07$ for $\Ray_S=10^{12}$, respectively. The autocorrelation function $R$ is also plotted versus the vertical separation in the viscous scale $(\delta z / L)\Ray_S^{1/3}$, see figure~\ref{fig:uzcor}(b). Interestingly, the two curves for $\Ray_S=10^{11}$ and $10^{12}$ collapse with each other, but they are different from the one for $\Ray_S=10^{10}$. Note that the three cases have the same density ratio $\Lambda=1.6$. This implies that when the distance between two plates is large enough and fingers cannot extend from one plate to the other, the vertical length of the salt fingers may be set by the density ratio.

We end this subsection by comparing our findings at high Rayleigh numbers to those reported in literatures. In~\cite{Stellmach2011}, the gravity-wave phase was followed by a spontaneous appearance of a layered phase and staircase-like scalar profiles. Such transition was not found in our simulation. The reason may be that $\Lew=100$ here is much larger than that in~\cite{Stellmach2011}. In our case the layered phase may occur at higher $\Ray_S$. More simulations are needed to clarify whether and when the layered phase can be realised in the current configuration. In the experiments of salt-sugar system wavy fingers were also observed both in the single finger layer occupying the entire tank or even in the finger layers bounded by two convection layers as in the staircase state~\citep{Krishnamurti2003}. However, they did not appear in the experiments of~\citet{Hage_Tilgner2010} and \citet{Kellner_Tilgner2014}. The exact conditions for the appearance of zonal flows and wavy fingers are not clear at this stage. The present Lewis number $\Lew=100$ is smaller than those in Tilgner's experiments ($\Lew\approx240$) but larger than those in~\cite{Krishnamurti2003} ($\Lew\approx3$), while the $\Ray_S$ for obtaining wavy fingers in our simulations is comparable to the highest $\Ray_S$ in Tilgner's experiments and much smaller than the $\Ray_S$ across the whole tank in~\cite{Krishnamurti2003}.

\section{Scaling laws for Hage \& Tilgner's experiments} 

For the scaling laws proposed in the previous two sections, one has to fit two exponents from the numerical results to collapse the data, i.e.~$\alpha^{\rm eff}_u$ for the Reynolds number $\Rey$ and $\alpha^{\rm eff}_T$ for the convective heat flux $\Nus_T-1$. The scaling laws for the flow structures, such as the finger width and the thickness of velocity boundary layers, can be deduced from the scaling of $\Rey$. In our simulations only one combination of the Prandtl numbers $(\Pra_T,\,\Pra_S)=(7,\,700)$ is considered. One may expect that the values of these exponents depend on the Prandtl numbers and should vary for different fluid system.

In order to test the scaling laws for the different fluid systems, we perform the previous analysis on the experimental data of~\citet{Hage_Tilgner2010} which have $(\Pra_T,\,\Pra_S)\approx(9,\,2200)$. Our previous study revealed that the GL prediction for the salinity transfer agrees with the experimental results~\citep{ddcjfm2015}. Here we further show that a complete description of the experimental results can be obtained by following the method described in the previous sections. In the experiments the heat flux was not measured, and we focus on the Reynolds number and the finger width. Recent studies revealed that when the density ratio is very small, the salt fingers will be replaced by large scale convection rolls and flow becomes very similar to the RB case~\citep{Kellner_Tilgner2014,ddcpnas16}. To confine ourselves to the salt-finger regime and be consistent with the current parameter range, we discard the data points with $\Lambda<0.1$ in the dataset of~\citet{Hage_Tilgner2010}.

\begin{figure}
\centering
\includegraphics[width=0.7\textwidth]{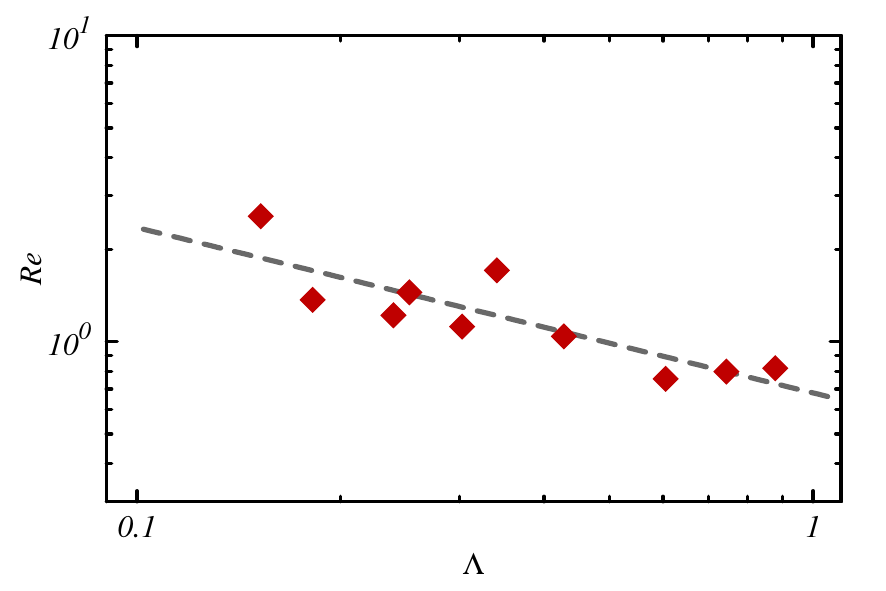}%
\caption{Dependences of $\Rey$ on $\Lambda$ for the experimental data from~\cite{Hage_Tilgner2010} with $\Ray_S\in(10^9,2\times10^9)$ and $\Lambda\ge0.1$. The dashed line indicates the linear regression giving $\Rey\sim\Lambda^{-\alpha^{\rm eff}_u}$ with $\alpha^{\rm eff}_u=0.54$.}
\label{fig:htalp}
\end{figure}

\begin{figure}
\centering
\includegraphics[width=\textwidth]{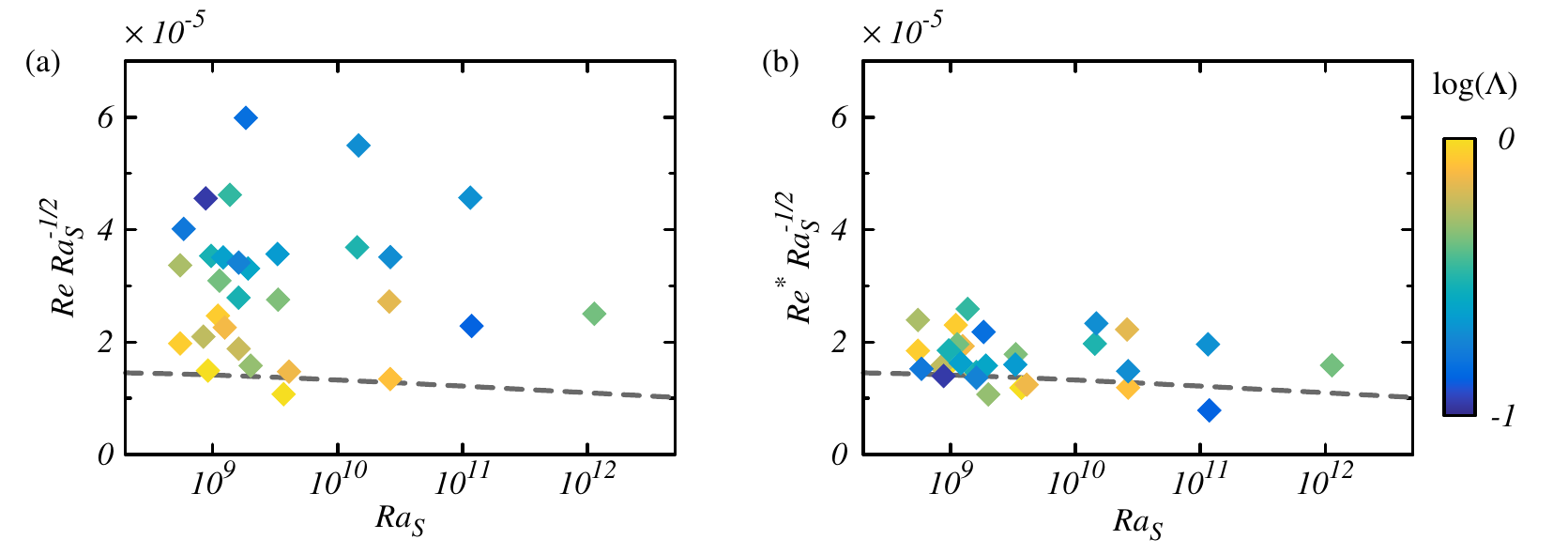}%
\caption{(a) Dependence of $\Rey$ versus $\Ray_S$ for the experimental data from~\cite{Hage_Tilgner2010} with $\Lambda\ge0.1$. (b) The same data of panel (a) but rescaled as $\Rey^*=\Rey\Lambda^{\alpha^{\rm eff}_u}$ with $\alpha^{\rm eff}_u=0.54$. Symbols are coloured according to the logarithm of $\Lambda$. The dashed lines are the predictions of the GL theory.}
\label{fig:htrey}
\end{figure}

To obtain the scaling laws for $\Rey$ and $d/L$, the only exponent we need to fit is $\alpha^{\rm eff}_u$. Since in the experiment it is very difficult to set the control parameters precisely, we choose ten cases within a narrow range $10^9<\Ray_S<2\times10^9$, see figure~\ref{fig:htalp}. By a linear regression we obtain $\Rey\sim\Lambda^{-\alpha^{\rm eff}_u}$ with $\alpha^{\rm eff}_u=0.54\pm0.10$. Similarly, we define the rescaled Reynolds number as $\Rey^*=\Rey\Lambda^{\alpha^{\rm eff}_u}$. In figure~\ref{fig:htrey} we plot both $\Rey$ and $\Rey^*$ against $\Ray_S$ for the experimental results with $\Lambda\ge0.1$ in a compensated form, and compare the results with the GL theory.\footnote{Here for the model coefficients we use the same values as given in~\citet{GLrefit2013}, and a transformation coefficient $\alpha=0.027$.} Indeed, compared to the original values the rescaled Reynolds number $\Rey^*$ collapses and is very close to the GL prediction.

\begin{figure}
\centering
\includegraphics[width=\textwidth]{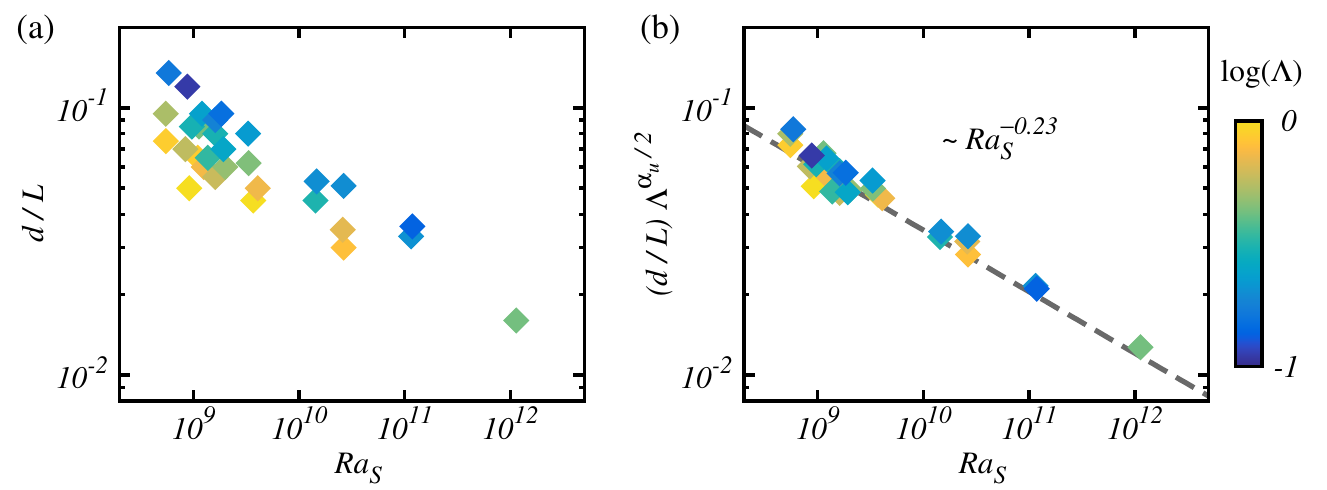}%
\caption{(a) The original finger width $d/L$ and (b) the rescaled finger width $(d/L)\Lambda^{\alpha^{\rm eff}_u/2}$ with $\alpha^{\rm eff}_u=0.54$ for the experimental data from~\cite{Hage_Tilgner2010} with $\Lambda\ge0.1$. Symbols are coloured according to the logarithm of $\Lambda$. In (b) the data points collapse after the rescaling, and the dashed line represents the linear regression with slope $-0.23$.}
\label{fig:htfgw}
\end{figure}

A rescaled finger width can be defined accordingly as $(d/L)\Lambda^{\alpha^{\rm eff}_u/2}$ with the $\alpha^{\rm eff}_u=0.54$. In figure~\ref{fig:htfgw} we plot the original and rescaled values of the finger width. The rescaled values collapse and its dependence on $\Ray_S$ follows a power-law scaling. The exponent obtained by a linear regression is $-0.23\pm0.04$. This value is very similar to that for $(\Pra_T,\,\Pra_S)=(7,\,700)$, i.e.~$-0.24\pm0.03$ as given in subsection 4.1. However, based on the current results it is not clear whether this exponent is universal for different Prandtl numbers.

It should be pointed out that our scaling relations are very close those given in~\citet{Hage_Tilgner2010}. \citet{Hage_Tilgner2010} proposed $\Rey\sim\Ray_T^{-1/2}\Ray_S\sim\Lambda^{-1/2}\Ray_S^{1/2}$. From figure~\ref{fig:htrey}(b) one observes that $\Rey^*\Ray_S^{-1/2}$ decreases very slowly as $\Ray_S$ increases. In other words, the current scaling law for the Reynolds number is roughly $\Rey\sim\Lambda^{-0.54}\Ray_S^{0.5}$. For the finger width, \citet{Hage_Tilgner2010} determined $d/L\sim\Ray_T^{-1/3}\Ray_S^{1/9}\sim\Lambda^{-1/3}\Ray_S^{-2/9}$, while our analysis gives $d/L\sim\Lambda^{-0.27}\Ray_S^{-0.23}$. Furthermore, \citet{Hage_Tilgner2010} suggested that $\Nus_S\sim\Ray_T^{-1/12}\Ray_S^{4/9}\sim\Lambda^{-1/12}\Ray_S^{13/36}$, i.e. a very weak dependence on $\Lambda$ and for $\Ray_S$ an exponent very close to $1/3$.

Finally, we want to stress that the effective exponents $\alpha^{\rm eff}_u$ and $\alpha^{\rm eff}_T$ take different values for different fluid system, i.e. they depend on two Prandtl numbers. In table~\ref{tab:expo} we summarise the values of $\alpha^{\rm eff}_u$ and $\alpha^{\rm eff}_T$ for three different combinations of $\Pra_T$ and $\Pra_S$, i.e. the current data, the experimental data from \citet{Hage_Tilgner2010}, and one group of data from \citet{ddcpnas16} with $(\Pra_T,~\Pra_S)=(7,~70)$, $\Ray_S=10^8$, and $0.1\le\Lambda\le1.0$. For the last dataset the finger regime occupies a smaller range of $\Lambda$ because $\Lew$ is smaller compared to the other two datasets. More simulations are needed, especially at different Prandtl numbers, to fully understand the physical origin of these exponents.  
\begin{table}
\renewcommand{\tabcolsep}{0.2cm}
\begin{center}
\begin{tabular}{c@{\hskip 1cm}ccc@{\hskip 1cm}cc}
 Source & $\Pra_T$ & $\Pra_S$ & $\Lew$ & $\alpha^{\rm eff}_u$ & $\alpha^{\rm eff}_T$ \\[0.2cm]
 Current dataset & 7 & 700 & 100 & $0.25\pm0.02$ & $0.75\pm0.03$  \\
 \citet{Hage_Tilgner2010} & 9 & 2200 & 244 & $0.54\pm0.10$ & ---   \\
 \citet{ddcpnas16} & 7 & 70 & 10 & $0.66\pm0.08$ & $0.54\pm0.03$  \\
\end{tabular}
\end{center}
\caption{The effective exponents $\alpha^{\rm eff}_u$ and $\alpha^{\rm eff}_T$ for different combinations of $\Pra_T$ and $\Pra_S$.}
\label{tab:expo}
\end{table}%

\section{Conclusions}

A systematic numerical study is carried out for the DDC flow bounded by two parallel plates at the Prandtl numbers $\Pra_T=7$ and $\Pra_S=700$, which are similar to the values of seawater. The salinity Rayleigh number $\Ray_S$ covers six decades in order of magnitude, i.e.~from $10^6$ to $10^{12}$, and the density ratio $\Lambda$ is between $0.1$ to $10$. The salinity Nusselt number $\Nus_S$ depends mainly on $\Ray_S$. The dependence of $\Nus_S$ on $\Ray_S$ is well captured by the GL theory with the same coefficients as determined for RB flow. For fixed $\Ray_S$, as the density ratio $\Lambda$ increases, $\Nus_S$ keeps constant, while both the Reynolds number $\Rey$ and the convective heat flux $\Nus_T-1$ decrease according to certain power-law scalings, namely $\Rey\sim\Lambda^{-\alpha^{\rm eff}_u}$ and $(\Nus_T-1)\sim\Lambda^{-\alpha^{\rm eff}_T}$. The two exponents are calculated from the numerical results, and for $(\Pra_T,\,\Pra_S)=(7,\,700)$ the values are $\alpha^{\rm eff}_u=0.25\pm0.02$ and $\alpha^{\rm eff}_T=0.75\pm0.03$. Then the rescaled Reynolds number and convective heat flux are introduced as $\Rey^*=\Rey\Lambda^{\alpha^{\rm eff}_u}$ and $\Nus_T^*=(\Nus_T-1)\Lambda^{\alpha^{\rm eff}_T}$. After rescaling, both $\Rey^*$ and $\Nus_T^*$ collapse for different $\Lambda$ and exhibit similar dependences on $\Ray_S$. We have interpreted this finding as the (stabilising) temperature field being advected by the velocity field like a passive scalar.

For fixed $\Ray_S$ and varying $\Lambda$, salt fingers have different horizontal size and velocity but transfer salinity with a similar rate. The flow fields on the horizontal mid plane indicate that salt fingers usually are circular in the horizontal sections. To generate a similar salinity flux at different density ratio, the horizontal width of salt fingers must scales as $d/L\sim\Rey^{1/2}\sim\Lambda^{-0.125}$ for fixed $\Ray_S$. Our numerical results confirm this argument, and the rescaled finger width $(d/L)\Lambda^{0.125}$ follows a single power-law scaling versus $\Ray_S$. In the current flow, the velocity boundary layer is driven by the vertical motion of salt fingers, and its thickness follows the same scaling laws as the finger width. The thickness of the salinity boundary layer, however, scales perfectly as $\lambda_s/L\sim\Nus^{-1}$, which is a natural result from the global balance between the salinity dissipation and Nusselt number, and the fact that the total salinity dissipation is dominated by the contribution from the boundary layers.

The scaling laws proposed for our numerical results are also tested against the experimental data of~\citet{Hage_Tilgner2010} with $(\Pra_T,\,\Pra_S)\approx(9,\,2200)$. The exponent $\alpha^{\rm eff}_u$ now has the value $0.54\pm0.10$. The rescaling of the Reynolds number and the finger width with this new $\alpha^{\rm eff}_u$ collapses the data points for all experimental cases with $\Lambda\ge0.1$. The dependence of the rescaled Reynolds number on $\Ray_S$ is very close to the GL prediction. The two effective exponents $\alpha^{\rm eff}_u$ and $\alpha^{\rm eff}_T$ are also calculated for an additional set of numerical results with $(\Pra_T,\,\Pra_S)=(7,\,70)$, $\Ray_S=10^8$, and $0.1\le\Lambda\le1.0$. Their values are $\alpha^{\rm eff}_u=0.66\pm0.08$ and $\alpha^{\rm eff}_T=0.54\pm0.03$, respectively. All these different values of $\alpha^{\rm eff}_u$ and $\alpha^{\rm eff}_T$ for different $(\Pra_T,\,\Pra_S)$ indicate that the two exponents depend on the properties of the fluid system, and more simulations are needed to fully understand their physical origins. 

When $\Ray_S$ is high enough, a stack of horizontal zonal flows emerge in the bulk region and have alternative flow directions. For the fluid system consider here, these zonal flow appears when $\Ray_S\ge10^{11}$. For these large $\Ray_S$ -- corresponding to big separation between two plates -- the salt fingers cannot extend from one plate to the opposite one, but instead have a vertical length which is smaller than the domain height. The inclined gravity waves, which were reported for the fully periodic simulations, were not observed in our bounded flows. This is probably due to the vertical constrain of the plates.

The current results propose various problems for future studies. The present scaling strategy should be further validated with other combinations of Prandtl numbers. Since the Prandtl numbers in our simulations are similar to those of seawater, it is of great interests to test the applicability of the current scaling relations to oceanic salt-finger layers. The same methodology may also be used to develop scaling relations for DDC in the diffusive regime, where the flow is driven by an unstable temperature difference and stabilized by a salinity difference. Finally, in our simulations even at $\Ray_S=10^{12}$, we did not observe the staircase state with alternating finger and convection layers. Determining the control parameters at which the staircase state will be realised in the current flow configuration would be very interesting and highly desired but requires more simulations.

\section*{Acknowledgements}

This study is supported by Foundation for Fundamental Research on Matter, and by the Netherlands Center for Multiscale Catalytic Energy Conversion (MCEC), an NWO Gravitation programme funded by the Ministry of Education, Culture and Science of the government of the Netherlands. The computing resources were provided by SURFsara, and through the PRACE project 2014112708.

\appendix
\section{Numerical details}

In the following tables we provide the details of our numerical simulations. 
\begin{table}
\begin{center}
\begin{tabular}{c@{\hskip -0.02cm}ccccccccccc}
  & $\Ray_S$  &  $\Lambda$  &  $\Nus_T$  &  $\Nus_S$  &  $\Rey$   &  $d/L$  &
   \begin{tabular}{c}$\lambda_u/L$ \\ \scalebox{0.8}{$(\times10^{-2})$}\end{tabular}  &  
   \begin{tabular}{c}$\lambda_s/L$ \\ \scalebox{0.8}{$(\times10^{-2})$}\end{tabular}  &
   $\Gamma$  &  $N_x \times N_z$  &  $n_x \times n_z$\\[0.2cm]
      & $1\times 10^{6}$  &   0.1  &   1.104  &   9.291  &   0.3903  &   0.7248  &    11.98  &    5.750  &     10  &  $240\times  120$  &  $5\times1$  \\ 
      & $1\times 10^{6}$  &     1  &   1.029  &   9.575  &   0.2647  &   0.4965  &    7.330  &    6.014  &     5  &  $192\times  96$  &  $3\times2$  \\ 
  $*$ & $1\times 10^{6}$  &    10  &   1.005  &   8.635  &   0.1392  &   0.4336  &    5.609  &    7.069  &     4  &  $192\times 120$  &  $2\times1$  \\ 
   & $2\times 10^{6}$  &   0.5  &   1.063  &   11.80  &   0.4070  &   0.4588  &    6.848  &    4.784  &     5  &  $256\times  96$  &  $3\times2$  \\ 
  $*$ & $2\times 10^{6}$  &     5  &   1.013  &   11.11  &   0.2301  &   0.3913  &    5.160  &    5.569  &     4  &  $192\times 120$  &  $2\times1$  \\ 
      & $5\times 10^{6}$  &   0.2  &   1.154  &   15.56  &   0.7092  &   0.3895  &    6.127  &    3.584  &     4  &  $256\times  96$  &  $3\times2$  \\ 
  $*$ & $5\times 10^{6}$  &     2  &   1.036  &   14.78  &   0.4284  &   0.3303  &    4.617  &    4.050  &     4  &  $192\times 120$  &  $3\times2$  \\ 
      & $8\times 10^{6}$  &  1.25  &   1.060  &   17.00  &   0.5800  &   0.2994  &    4.406  &    3.451  &     4  &  $240\times 144$  &  $3\times2$  \\ 
      & $1\times 10^{7}$  &   0.1  &   1.368  &   18.39  &    1.112  &   0.4112  &    6.314  &    2.971  &     4  &  $256\times 144$  &  $4\times2$  \\ 
      & $1\times 10^{7}$  &   0.4  &   1.127  &   18.79  &   0.8295  &   0.2942  &    4.571  &    3.021  &     2  &  $192\times 120$  &  $3\times2$  \\ 
  $*$ & $1\times 10^{7}$  &     1  &   1.078  &   18.23  &   0.6723  &   0.2964  &    4.317  &    3.205  &     4  &  $240\times 144$  &  $3\times2$  \\ 
      & $1\times 10^{7}$  &     2  &   1.045  &   18.33  &   0.5770  &   0.2729  &    3.795  &    3.244  &     2  &  $144\times 120$  &  $3\times2$  \\ 
      & $1\times 10^{7}$  &     4  &   1.026  &   18.14  &   0.4820  &   0.2528  &    3.417  &    3.349  &     2  &  $144\times 120$  &  $3\times2$  \\ 
  $*$ & $1\times 10^{7}$  &    10  &   1.012  &   17.44  &   0.3591  &   0.2402  &    2.978  &    3.399  &     2  &  $192\times 144$  &  $2\times2$  \\ 
  $*$ & $2\times 10^{7}$  &   0.5  &   1.169  &   22.09  &    1.024  &   0.2726  &    4.164  &    2.592  &   1.6  &  $192\times 144$  &  $3\times2$  \\ 
  $*$ & $2\times 10^{7}$  &     5  &   1.028  &   22.43  &   0.5963  &   0.2173  &    2.756  &    2.732  &     2  &  $192\times 144$  &  $2\times2$  \\ 
  $*$ & $5\times 10^{7}$  &   0.2  &   1.443  &   29.01  &    1.830  &   0.2271  &    3.963  &    1.911  &   1.6  &  $240\times 192$  &  $3\times2$  \\ 
  $*$ & $5\times 10^{7}$  &     2  &   1.082  &   29.03  &    1.088  &   0.1960  &    2.595  &    2.094  &     2  &  $240\times 192$  &  $3\times2$  \\ 
      & $8\times 10^{7}$  &  1.25  &   1.133  &   34.01  &    1.526  &   0.1709  &    2.379  &    1.761  &     2  &  $288\times 216$  &  $3\times2$  \\ 
\end{tabular}
\end{center}
\caption{Summary of the control parameters and the numerical results. For all simulations $\Pra_T=7$ and $\Pra_S=700$. Columns from left to right: the salinity Rayleigh numbers $\Ray_S$, the density ratio $\Lambda$, the heat and salinity Nusselt numbers $\Nus_T$ and $\Nus_S$, the Reynolds number $\Rey$, the finger width $d/L$, the thicknesses of velocity and salinity boundary layers $\lambda_u$ and $\lambda_s$, the aspect ratio $\Gamma$ of the computational domain, the base resolutions and refinement factors in the $x$ and $z$ directions, respectively. The domain size and resolution in the $y$-direction are the same as those in the $x$-direction. Asterisks mark the cases from~\cite{ddcjfm2015}, in which the Reynolds number was consistently underestimated by around $20\%$ due to a round-off error introduced in the original post processing analysis. Here all these values were corrected.}
\label{tab:nume}
\end{table}%

\begin{table*}
\begin{center}
\begin{tabular}{c@{\hskip -0.02cm}ccccccccccc}
  & $\Ray_S$  &  $\Lambda$  &  $\Nus_T$  &  $\Nus_S$  &  $\Rey$   &  $d/L$  &
   \begin{tabular}{c}$\lambda_u/L$ \\ \scalebox{0.8}{$(\times10^{-2})$}\end{tabular}  &  
   \begin{tabular}{c}$\lambda_s/L$ \\ \scalebox{0.8}{$(\times10^{-2})$}\end{tabular}  &
   $\Gamma$  &  $N_x \times N_z$  &  $n_x \times n_z$\\[0.2cm] 
  $*$ & $1\times 10^{8}$  &   0.1  &   1.890  &   35.46  &    2.936  &   0.2177  &    3.676  &    1.551  &   1.6  &  $288\times 240$  &  $3\times2$  \\ 
      & $1\times 10^{8}$  &   0.4  &   1.318  &   36.66  &    2.182  &   0.1734  &    2.691  &    1.582  &   1.6  &  $288\times 240$  &  $3\times2$  \\ 
  $*$ & $1\times 10^{8}$  &     1  &   1.173  &   36.19  &    1.755  &   0.1692  &    2.361  &    1.651  &     2  &  $288\times 216$  &  $3\times2$  \\ 
      & $1\times 10^{8}$  &     2  &   1.100  &   36.05  &    1.459  &   0.1581  &    2.133  &    1.681  &   1.6  &  $240\times 216$  &  $3\times2$  \\ 
      & $1\times 10^{8}$  &     4  &   1.058  &   36.48  &    1.225  &   0.1478  &    1.871  &    1.672  &   1.6  &  $240\times 216$  &  $3\times2$  \\ 
      & $1\times 10^{8}$  &    10  &   1.026  &   34.84  &   0.8992  &   0.1362  &    1.631  &    1.700  &   1.2  &  $216\times 192$  &  $2\times2$  \\ 
  $*$ & $2\times 10^{8}$  &   0.5  &   1.396  &   42.87  &    2.646  &   0.1591  &    2.387  &    1.360  &     1  &  $240\times 192$  &  $3\times3$  \\ 
      & $2\times 10^{8}$  &     5  &   1.059  &   44.97  &    1.539  &   0.1166  &    1.488  &    1.330  &   1.2  &  $240\times 192$  &  $3\times3$  \\ 
  $*$ & $5\times 10^{8}$  &   0.2  &   2.026  &   56.60  &    4.869  &   0.1285  &    2.295  &   0.9862  &     1  &  $240\times 288$  &  $4\times3$  \\ 
      & $5\times 10^{8}$  &     2  &   1.168  &   58.90  &    2.853  &   0.1043  &    1.389  &    1.022  &   1.2  &  $288\times 288$  &  $3\times2$  \\ 
      & $8\times 10^{8}$  &  1.25  &   1.287  &   67.32  &    3.891  &   0.0986  &    1.335  &   0.9001  &   1.2  &  $360\times 384$  &  $3\times2$  \\ 
  $*$ & $1\times 10^{9}$  &   0.1  &   3.103  &   68.40  &    7.752  &   0.1142  &    2.302  &   0.7922  &   0.8  &  $288\times 288$  &  $3\times3$  \\ 
      & $1\times 10^{9}$  &     1  &   1.362  &   72.34  &    4.539  &   0.0932  &    1.303  &   0.8263  &   1.2  &  $288\times 288$  &  $4\times3$  \\ 
      & $1\times 10^{9}$  &  1.05  &   1.352  &   71.89  &    4.446  &   0.0938  &    1.322  &   0.8338  &   1.2  &  $384\times 384$  &  $4\times3$  \\ 
      & $1\times 10^{9}$  &   1.2  &   1.320  &   72.23  &    4.290  &   0.0945  &    1.286  &   0.8238  &     1  &  $256\times 384$  &  $4\times3$  \\ 
      & $1\times 10^{9}$  &   1.6  &   1.255  &   71.66  &    3.941  &   0.0934  &    1.243  &   0.8350  &   0.7  &  $256\times 384$  &  $3\times3$  \\ 
      & $1\times 10^{9}$  &     2  &   1.213  &   72.33  &    3.740  &   0.0917  &    1.184  &   0.8236  &   0.6  &  $256\times 384$  &  $3\times3$  \\ 
      & $1\times 10^{9}$  &    10  &   1.052  &   70.30  &    2.347  &   0.0768  &   0.8772  &   0.8269  &   0.8  &  $240\times 240$  &  $3\times3$  \\ 
      & $1\times 10^{10}$  &  1.05  &   1.697  &   144.3  &    11.01  &   0.0583  &   0.7615  &   0.3810  &   0.8  &  $576\times 768$  &  $3\times3$  \\ 
      & $1\times 10^{10}$  &   1.2  &   1.651  &   146.5  &    10.82  &   0.0591  &   0.7468  &   0.3794  &   0.6  &  $384\times 768$  &  $4\times3$  \\ 
      & $1\times 10^{10}$  &   1.6  &   1.530  &   145.4  &    10.11  &   0.0548  &   0.7016  &   0.4016  &   0.5  &  $384\times 768$  &  $3\times3$  \\ 
      & $1\times 10^{10}$  &     2  &   1.454  &   147.8  &    9.694  &   0.0513  &   0.6462  &   0.4087  &   0.4  &  $256\times 512$  &  $4\times4$  \\ 
      & $1\times 10^{11}$  &   1.6  &   1.920  &   297.7  &    26.12  &   0.0294  &   0.3834  &   0.1755  &  0.32  &  $480\times 960$  &  $3\times4$  \\ 
      & $1\times 10^{12}$  &   1.6  &   2.677  &   599.6  &    66.60  &   0.0152  &   0.2093  &  0.08132  &  0.16  &  $768\times2048$  &  $2\times4$  \\
\end{tabular}
\end{center}
\caption{Continue of table~\ref{tab:nume}.}
\end{table*}%

\newpage

\end{document}